\documentclass[twocolumn]{openjournal}

\usepackage[dvipsnames]{xcolor}

\usepackage{graphicx}
\usepackage{xcolor}
\newcommand{\Gaia}{{\it Gaia}}
\newcommand{\feh}{\mbox{[Fe/H]}}
\newcommand{\mh}{\mbox{[M/H]}}
\newcommand{\teff}{\ensuremath{T_{\rm eff}}}
\newcommand{\logg}{\ensuremath{\log g}}
\newcommand{\vmic}{\ensuremath{v_{\rm mic}}}
\newcommand{\kms}{\ensuremath{\,{\rm km\,s^{-1}}}}
\newcommand{\Korg}{\texttt{Korg}}
\newcommand{\rev}[1]{#1}
\renewcommand{\facility}[1]{\par\noindent\textit{Facilities:} #1}
\providecommand{\software}[1]{\par\noindent\textit{Software:} #1}
\makeatletter
\def\frontmatter@title@below{\vspace*{0.1in}}
\def\frontmatter@above@affilgroup{\vspace*{0.08in}}
\makeatother

\definecolor{Steel}{RGB}{112,128,144}
\definecolor{MetallicSilver}{RGB}{192,192,192}
\definecolor{Metal}{RGB}{170,176,184}

\definecolor{MetallicBronze}{RGB}{205,127,50}

\shorttitle{PANTERA. I.}
\shortauthors{Saad et al.}

\begin{document}

\title{PANTERA. I. Hunting for the Most Metal-Rich Stars in the Solar Neighborhood with High-Resolution Spectroscopy}

\author{Serat M. Saad\altaffilmark{1}}
\altaffiltext{1}{saad.104@osu.edu}
\affiliation{Department of Astronomy, The Ohio State University, 140 West 18th Avenue,
Columbus, OH 43210, USA}

\author{D. M. Rowan\altaffilmark{2}}
\altaffiltext{2}{NHFP Hubble Fellow}
\affiliation{Department of Astronomy, University of California, Berkeley, CA 94720, USA}

\author{K. Z. Stanek}
\affiliation{Department of Astronomy, The Ohio State University, 140 West 18th Avenue,
Columbus, OH 43210, USA}
\affiliation{Center for Cosmology and Astroparticle Physics, The Ohio State University, 191 W. Woodruff Avenue, Columbus, OH, 43210, USA}

\author{Ilya Ilyin}
\affiliation{Leibniz-Institut for Astrophysics Potsdam (AIP), An der Sternwarte 16, D14482 Potsdam, Germany}

\author{Benjamin J. Fulton}
\affiliation{Department of Astronomy, California Institute of Technology, Pasadena, CA 91125, USA}

\author{Howard Isaacson}
\affiliation{Department of Astronomy, University of California, Berkeley, CA 94720, USA}

\author{Jessica R. Lu}
\affiliation{Department of Astronomy, University of California, Berkeley, CA 94720, USA}
\affiliation{Space Sciences Laboratory, University of California, Berkeley, CA 94720, USA}

\begin{abstract}

We present PANTERA (Project for Astrophysical Nucleosynthesis and Targeted Exploration of metal-Rich Abundances), a high-resolution spectroscopic survey of the most metal-rich stars in the solar neighborhood. In this first paper, we report iron abundances for 56 metal-rich stars, selected from \Gaia\ DR3 XP spectrophotometric metallicities. These targets were observed with the PEPSI spectrograph on the Large Binocular Telescope (LBT), the Levy spectrograph on the Automated Planet Finder (APF) telescope, and the HIRES spectrograph on the Keck telescope. We \rev{derive} \feh\ from an equivalent-width analysis of iron lines, with the effective temperature taken from photometry and the surface gravity from the \Gaia\ parallax. We verify our measurement using \Gaia\ benchmark stars, observing some sources with more than one spectrograph, using a second synthesis code, and an independent equivalent-width measurement. We find that the \Gaia-XP metallicities over-predict \feh\ for part of our sample: they \rev{agree with our values} for the dwarfs, though a selection bias limits what \rev{this} agreement can show, and \rev{exceed our \feh\ by} \rev{$0.15\pm0.02$}~dex \rev{on average} for the cool giants. We attribute the over-prediction in part to the strong blue line blanketing of the cool metal-rich giants and to the survey labels on which the XP metallicities were trained. We \rev{find that} 25 of the 56 stars \rev{have} $\feh > \rev{+}0.4$, with the most iron-rich stars reaching $\feh=+0.58$. We discuss the implications of our result on studying metal-rich populations.

\end{abstract}

\keywords{Stellar abundances, Metallicity, Galaxy chemical evolution,
High resolution spectroscopy}

\section{Introduction}\label{sec:intro}
Galactic chemical evolution models suggest that the metallicity of the interstellar medium near the Sun has changed little over the past few Gyr \citep{Chiappini03}. Stars with metallicities well above the solar value are therefore not expected to form in the Solar neighborhood. Instead, their presence near the Sun is usually attributed to radial migration from the inner Galaxy, where the interstellar medium reaches higher metallicities \citep{Sellwood02,Roskar08,Schonrich09,Minchev10}. \rev{Stars well above the solar metallicity, often called super-metal-rich (SMR) stars, were identified in the Solar neighborhood several decades ago \citep{Grenon89,Castro97}, and a few reach $\feh\simeq+0.4$ or more \citep{Castro97,Trevisan11}. These stars are thought to be} tracers of radial migration and disk heating \citep{Frankel18,Kordopatis15,Kordopatis25}, influence of the Galactic bar \citep{Bovy19,Nepal24}, and gas inflow and outflow \citep{Grand19}.

Recent work with large spectroscopic and astrometric surveys finds that the SMR stars are concentrated toward the inner disk and the Galactic bar \citep{Queiroz21,Chen22}, while studies using main-sequence turnoff and subgiant stars from the \Gaia\ radial velocities find that these SMR stars are old, with ages above $7$~Gyr with a bimodal distribution in guiding radius \citep{Nepal24}. Other surveys of the disk metallicity distribution of Milky Way find a metal-rich tail that is well populated only toward the inner Galaxy \citep{Hayden15,Bensby14}. Near-infrared spectroscopy of the nuclear star cluster and the nuclear stellar disk at the Galactic center also finds a mostly metal-rich population \citep{FeldmeierKrause17,Schultheis21,Ryde25,Nandakumar25}, with one star above $\mh=+0.6$ \citep{Do18}. However, \citet{Chen19} use \rev{Large Sky Area Multi-Object Fiber Spectroscopic Telescope (LAMOST)} spectra to identify SMR stars and find that only about half of the sample is kinematically consistent with migration from the inner Galaxy.

Metal-rich stars are also calibrators for stellar models. At super-solar metallicity the enhanced opacity changes the energy transport, and the mass loss on the red giant branch is expected to increase \citep{Reimers75}. The old, metal-rich open cluster NGC~6791, with $\feh\sim+0.4$ and an age near $8$~Gyr \citep{Brogaard12}, hosts a population of low-mass helium white dwarfs that can be explained by enhanced mass loss on the giant branch \citep{Hansen05,Kalirai07}. The physics of red giant mass loss is not well understood, and simple empirical relations are often used in its place. The empirical evidence is mixed: some cluster studies find mass loss that increases with metallicity \citep{Tailo21}, while asteroseismic mass measurements in old clusters point to a weaker dependence \citep{Miglio12, Li2025, Roberts2026}. Thus, a sample of giants at extreme metallicity gives a direct test of these models.

On smaller scales, the properties of planets correlate with the metallicity of the host star. The frequency of giant planets rises with the host metallicity \citep{Fischer05}, and the occurrence of short-period planets also increases toward higher metallicity \citep{Mulders16}, which supports core-accretion models of planet formation \citep{Ida04}. Giant planets are more common around the most metal-rich stars \citep{Petigura18}, yet the planet-metallicity relation is not well constrained above $\feh\sim+0.3$, where few confirmed hosts are known \citep{Adibekyan12}. The most metal-rich stars in the solar neighborhood are therefore a useful target list for future radial-velocity monitoring.

Previous high-resolution studies of SMR stars \citep{Castro97,Feltzing01,Trevisan11,Dantas23} show that these populations are chemically diverse, with several subgroups and ages from $7$ to $10$~Gyr. These studies used fewer than about $70$ stars selected from pre-\Gaia\ catalogs, and none targeted the high-metallicity tail above $\feh=+0.4$ with a \Gaia-based selection. \Gaia\ \rev{Data Release 3 (DR3)} now provides metallicities from \rev{the low-resolution blue and red photometer (BP/RP, or XP) spectra} for $\sim175$ million stars \citep{Andrae23} and for $\sim220$ million stars in an independent analysis \citep{Zhang23}, with a typical uncertainty near $0.1$~dex in \mh, alongside metallicities from the \rev{Radial Velocity Spectrometer (RVS)} \citep{RecioBlanco23}, which together make a large, uniform selection of metal-rich candidates possible. The XP metallicities are low-resolution and data-driven, trained on external labels whose metal-rich calibrators are few, so a direct test with high-resolution spectroscopy is needed before the XP metal-rich candidates can be trusted for studying the inner disk.

We aim to change that with PANTERA, the Project for Astrophysical Nucleosynthesis and Targeted Exploration of metal-Rich Abundances. PANTERA is a program to survey and understand the most metal-rich stars in the solar neighborhood. In this first paper we begin with the \Gaia-XP candidates. We select metal-rich candidates from the \Gaia-XP metallicities and observe them at high spectral resolution to measure accurate iron abundances and to test the XP scale. This first paper presents the method and the iron abundances for 56 stars observed with three spectrographs. We describe the sample selection and the observations in Section~\ref{sec:data}, the abundance analysis and its validation in Section~\ref{sec:analysis}, the iron abundances and their comparison with the \Gaia-XP scale in Section~\ref{sec:results}, and the implications in Section~\ref{sec:disc}. We summarize in Section~\ref{sec:conclusions}.

\section{Target Selection and Observations}\label{sec:data}

\begin{deluxetable*}{lcccccccc}
\tabletypesize{\footnotesize}
\tablecaption{PANTERA targets: positions, Galactic coordinates, \Gaia\ distances, magnitudes, \Gaia-XP metallicities, and instruments. A representative subset is shown here; the full table for all 56 stars is published in the machine-readable form.\label{tab:targets}}
\tablehead{\colhead{\Gaia\ DR3 source\_id} & \colhead{R.A.} & \colhead{Decl.} & \colhead{$l$} & \colhead{$b$} & \colhead{$d$} & \colhead{$G$} & \colhead{\mh$_{\rm XP}$} & \colhead{Instrument} \\ \colhead{} & \colhead{(deg)} & \colhead{(deg)} & \colhead{(deg)} & \colhead{(deg)} & \colhead{(pc)} & \colhead{(mag)} & \colhead{(dex)} & \colhead{}}
\startdata
4547137062312951680 & 261.24032 & +16.93493 & 39.25 & +26.52 & 221 & 7.40 & $+0.41$ & HIRES+PEPSI \\
1291436403222476928 & 226.53371 & +34.75342 & 56.16 & +60.28 & 229 & 7.52 & $+0.42$ & APF+HIRES \\
2207311069265753088 & 344.27742 & +63.03213 & 110.38 & +2.98 & 60 & 8.35 & $+0.44$ & HIRES \\
1810432623428484992 & 303.70796 & +19.16331 & 59.71 & -8.60 & 459 & 10.03 & $+0.41$ & APF \\
4430730494968490368 & 234.06402 & +7.08052 & 13.49 & +46.11 & 120 & 10.18 & $+0.43$ & PEPSI \\
1315367857917145856 & 241.52926 & +25.64117 & 42.46 & +46.65 & 215 & 10.22 & $+0.43$ & PEPSI \\
1972949478614475520 & 325.38300 & +43.39273 & 90.27 & -7.11 & 230 & 10.30 & $+0.41$ & HIRES \\
2222930319131123584 & 325.97451 & +68.41728 & 107.03 & +11.55 & 144 & 10.40 & $+0.41$ & HIRES \\
2241587416547315584 & 297.21581 & +64.35435 & 96.71 & +18.39 & 309 & 10.48 & $+0.41$ & APF \\
1332825044549520640 & 246.98675 & +40.03574 & 63.60 & +43.84 & 624 & 10.55 & $+0.42$ & HIRES \\
2076073979856273664 & 300.53118 & +44.14328 & 79.38 & +7.11 & 267 & 10.70 & $+0.44$ & PEPSI \\
4591824513402826112 & 274.72797 & +31.58410 & 58.93 & +20.25 & 951 & 10.84 & $+0.39$ & APF \\
\enddata
\tablecomments{Galactic coordinates are from \Gaia\ DR3. The distance $d$ is the photogeometric estimate of \citet{Bailer-Jones21}. This table is published in its entirety, for all $56$ stars, at \rev{\url{https://github.com/panterasurvey/pantera1}}. A portion is shown here for guidance regarding its form and content.}
\end{deluxetable*}

\begin{figure}
\centering
\includegraphics[width=\columnwidth]{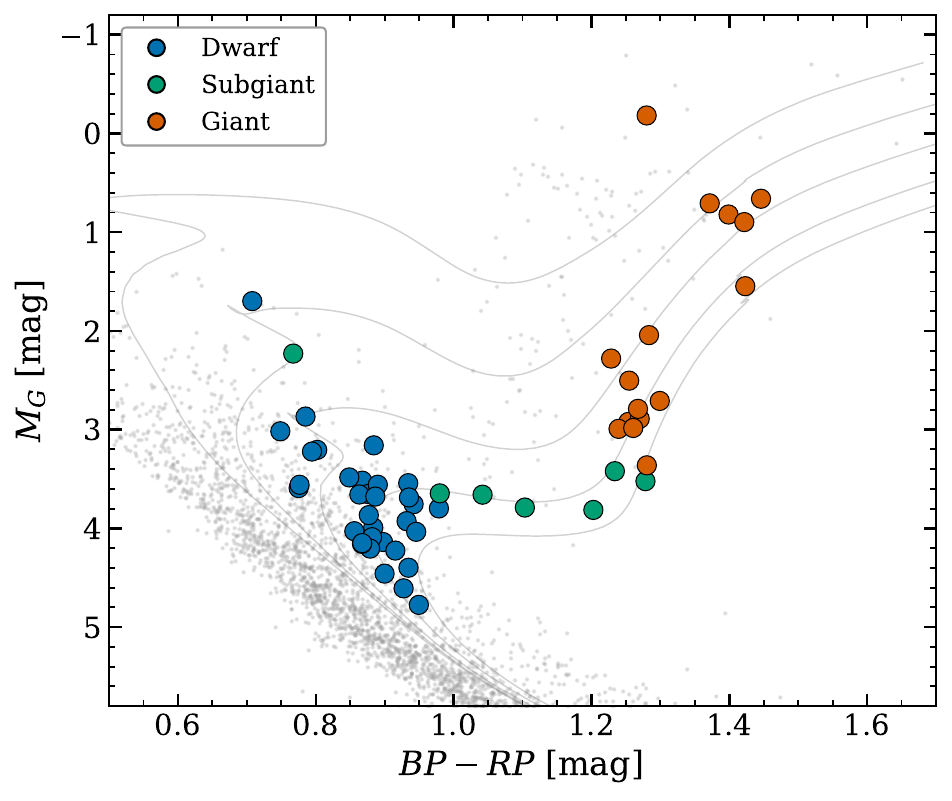}
\caption{\Gaia\ color-magnitude diagram ($M_G$ vs. $BP-RP$) of the 56 observed targets, colored by evolutionary type, on a background of a random nearby \Gaia\ field sample (grey). The targets span the warm dwarfs, the subgiant branch, and the red giant branch. As expected, the metal-rich targets appear redder than the Solar-metallicity population. \rev{Grey curves are MIST isochrones at $\feh=+0.4$, the median of our sample, for ages of $1$, $2$, $4$, $8$, and $12.6$~Gyr, which we use to classify the stars (Section~\ref{sec:params}).}}
\label{fig:cmd}
\end{figure}

\subsection{Target Selection}\label{sec:select}
We select metal-rich candidate stars from the catalog of \Gaia-XP metallicities of \citet{Andrae23}\rev{, which are derived with the extreme gradient boosting (XGBoost) algorithm}. Because both the XP metallicity and the accessible survey volume depend on luminosity, we apply the metallicity and distance criteria by evolutionary state. For dwarfs and subgiants we require an XGBoost XP metallicity $\mh_{\rm XP}>+0.4$ and a distance $<500$~pc. For giants (XGBoost $\log g<3.5$) we lower the threshold to $\mh_{\rm XP}>+0.35$ and impose no distance limit\rev{, because for these luminous stars our $G<13$ limit, rather than a distance cut, sets the survey volume. The most distant giant we observed lies at $1.6$~kpc.} Both samples share the magnitude and astrometric-quality cuts $G<13$, parallax signal-to-noise $>5$, $\delta>-13^\circ$, \rev{renormalized unit weight error} ${\rm RUWE}<1.4$, and \texttt{ipd\_frac\_multi\_peak}$\leq2$, the last two to suppress blended sources and close binaries that would contaminate the XP spectrum. In practice we join the \citet{Andrae23} XGBoost table to the \Gaia\ DR3 source catalog and keep the sources that pass these metallicity, distance, magnitude, and astrometric-quality cuts. The parent catalog provides XGBoost metallicities for about $175$ million stars with XP spectra. The metallicity thresholds remove almost all of them, and the magnitude, distance, declination, and astrometric-quality cuts reduce the remainder to about $3100$ candidates. The final step from the candidate list to the observed sample is set by telescope access. The exposure times needed for our signal-to-noise target favor the brightest candidates, and the observing semesters set the accessible part of the sky. \rev{We provide the ADQL queries for the two selections with our code (see Data Availability).}

Figure~\ref{fig:cmd} shows the color-magnitude diagram of the 56 observed targets on a \Gaia\ field background. They span the giant, subgiant, and dwarf sequences. The 56 observed targets are the brightest high-metallicity targets that were observable in the 2026 semesters and span \teff\ from $4435$ to $6236$~K and \logg\ from $2.2$ to $4.5$, so the sample runs from cool giants \rev{(all with $\teff<5000$~K)} through subgiants to warm dwarfs \rev{(all with $\teff>5400$~K)}. \rev{We assign the evolutionary types from isochrones (Section~\ref{sec:params}).} The positions, magnitudes, \Gaia-XP metallicities, and instruments of the targets are listed in Table~\ref{tab:targets}.

\subsection{Spectroscopic Observations}\label{sec:obs}
We observed the targets with three high-resolution spectrographs. We observed 31 stars with the Potsdam Echelle Polarimetric and Spectroscopic Instrument (PEPSI; \citealt{Strassmeier15}) on the Large Binocular Telescope (LBT). We used the $200\,\mu$m fiber, which gives a resolving power of about $R=120{,}000$, with the cross-disperser pairs that cover the optical range from about $4800$ to $7400$~\AA. The large fiber diameter allows a signal-to-noise $>100$ in exposures of a few hundred to about $1200$~s. We observed 14 stars with the Levy slit-fed optical echelle spectrograph on the $2.4$-m Automated Planet Finder (APF) telescope at Lick Observatory \citep{Vanderburg16}, at a resolving power of $\sim110{,}000$ over the $3740$--$9700$~\AA\ range using the $1\arcsec\times3\arcsec$ Decker-W slit and exposure times of 1800s. We observed 17 stars with the \rev{High Resolution Echelle Spectrometer (HIRES;} \citealt{Vogt94}\rev{)} on the Keck~I telescope, at a resolving power of $\sim50{,}000$ over $\sim3600$--$8000$~\AA{} with the C2 decker in the standard California Planet Search setup \citep{Howard10} with typical exposure times of 400--800s. The APF and HIRES spectra were taken without the iodine in the beam. Six of the stars were observed with multiple spectrographs, which gives 62 spectra of 56 distinct stars and six pairs for a cross-instrument test (Section~\ref{sec:valid}). The targets were selected to reach a signal-to-noise $\sim75$ per resolution element in the region used for the iron lines. The achieved values are higher. Measured from the scatter about the continuum near $6150$~\AA, the median signal-to-noise per resolution element is about $300$ for the PEPSI spectra, $140$ for the APF spectra, and $190$ for the HIRES spectra, with the lowest spectrum near the selection target of $75$.

\section{Analysis}
\label{sec:analysis}

We \rev{derive} the iron abundance (\feh) of each star with a classical equivalent-width analysis \rev{at fixed stellar parameters. We take \teff\ from photometry and \logg\ from the \Gaia\ parallax, and we set only the microturbulence from the spectrum, by requiring that the \ion{Fe}{1} abundances show no trend with reduced equivalent width. We use the excitation and ionization balance of the iron lines only to check the adopted parameters (Section~\ref{sec:paramcheck}). We apply the same analysis to every spectrum, in six steps}: (i) spectral preparation, (ii) adoption of the stellar parameters, (iii) construction of the line list, (iv) measurement of the equivalent widths, (v) conversion of the equivalent widths to abundances with the microturbulence set by the reduced-equivalent-width balance, and (vi) an error estimate. We describe each step below and then describe the validation tests (Section~\ref{sec:valid}).

\begin{figure*}
\centering
\includegraphics[width=\textwidth]{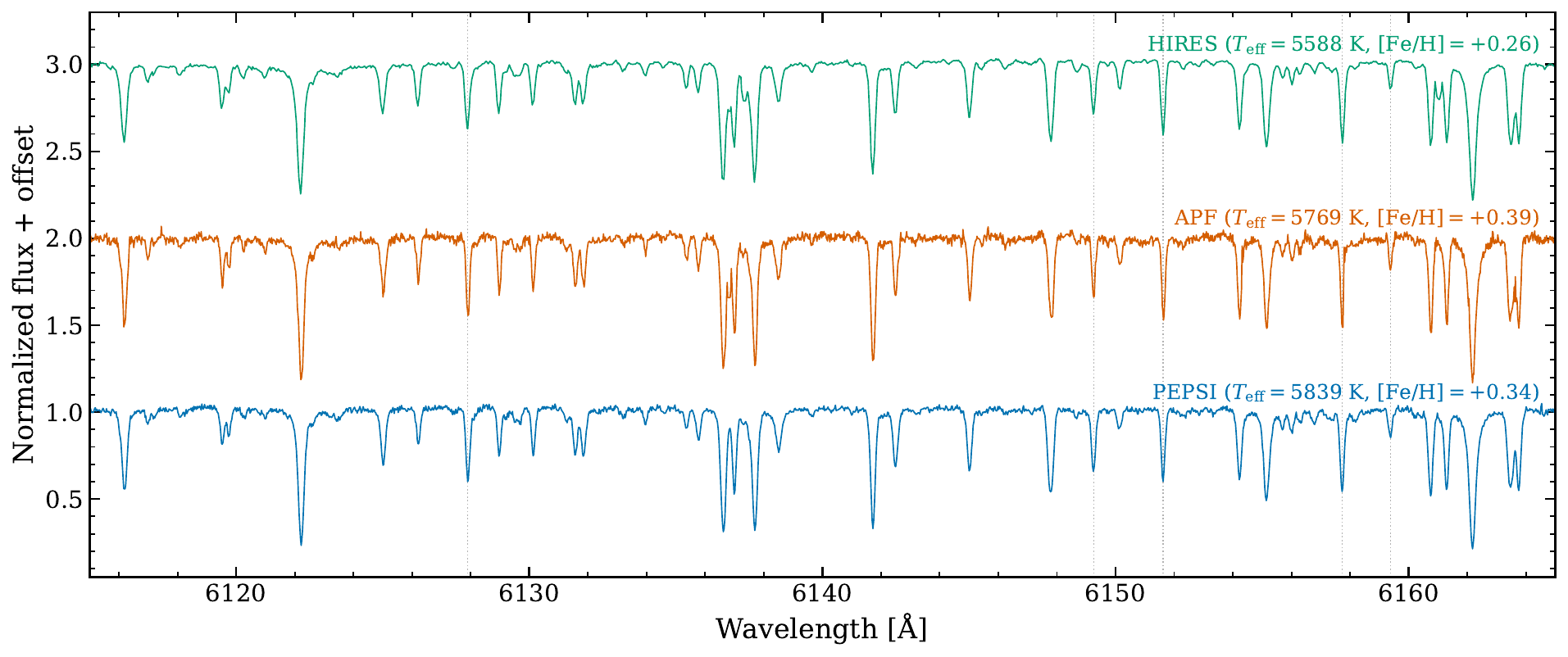}
\caption{Three representative warm metal-rich dwarfs, one observed with each of the three spectrographs (PEPSI, APF, HIRES), in a $50$~\AA\ window rich in iron lines, offset vertically for clarity. The three stars have similar stellar parameters, listed for each spectrum, and the iron lines from our line list (dotted) line up across the instruments, which shows that the three reductions place the same lines on a common scale.}
\label{fig:spectra}
\end{figure*}

\subsection{Spectral preparation}\label{sec:prep}
The PEPSI spectra are reduced with the Spectroscopic Data Systems pipeline \citep{Strassmeier18}. We then shift the spectrum to the rest frame, mask the telluric lines, and apply a local continuum normalization \citep{BlancoCuaresma14}. The APF and HIRES spectra are reduced to one-dimensional, wavelength-calibrated echelle orders with the standard California Planet Search pipeline \citep{Howard10, Fulton15}. For each of these we normalize every order by its upper envelope, stitch the orders onto a common wavelength grid by inverse-variance weighting, and shift each spectrum to the rest frame from a cross-correlation against an \ion{Fe}{1} line mask. Figure~\ref{fig:spectra} shows the same iron-line region for three similar warm metal-rich dwarfs, one from each spectrograph. We found the iron lines fall at matching positions across the instruments.

\rev{The PEPSI pipeline applies the barycentric correction, and for the 31 PEPSI stars the pipeline radial velocities agree with the \Gaia\ DR3 values to a median of $+0.26\kms$ with a scatter of $0.35\kms$, and no star differs by more than $1.4\kms$. For the APF and HIRES spectra we apply no barycentric correction, because we use the cross-correlation only to shift each spectrum to the rest frame, so we do not use these shifts as radial velocities.}

\begin{figure}
\centering
\includegraphics[width=\columnwidth]{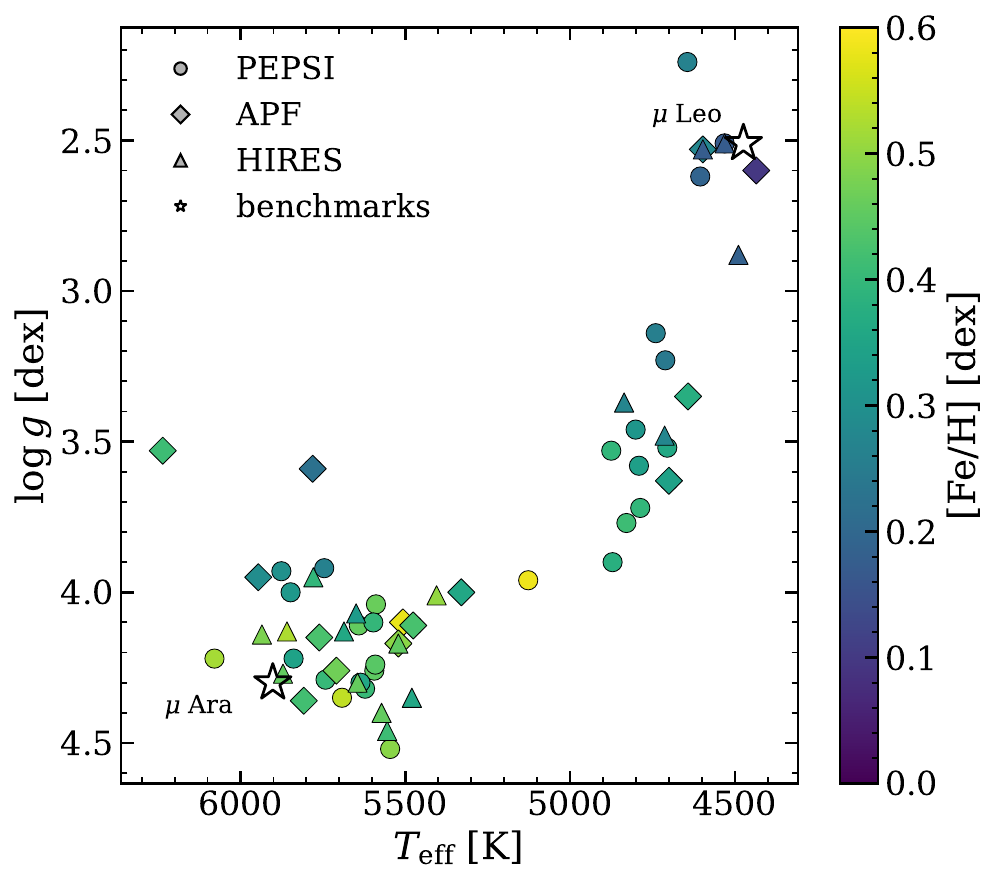}
\caption{The 56 PANTERA targets in the $\teff$--$\logg$ plane, colored by the derived \feh\ and marked by instrument. The two \Gaia\ FGK benchmark stars used for validation ($\mu$~Leo, $\mu$~Ara; stars) span the parameter range of the sample, which runs from cool giants to warm dwarfs.}
\label{fig:kiel}
\end{figure}

\subsection{Stellar parameters}\label{sec:params}
\rev{We set \teff\ and \logg\ from photometry and the \Gaia\ parallax and keep both fixed during the analysis. We avoid a fully spectroscopic solution because in these strongly blanketed spectra \teff, \logg, metallicity, and \vmic\ are degenerate, and the \ion{Fe}{1}/\ion{Fe}{2} ionization balance is biased by the over-ionization of \ion{Fe}{1} that is missed under local thermodynamic equilibrium (LTE) \citep{Bergemann12,Lind12}. Our external parameters are also independent of the \Gaia-XP scale that we test in Section~\ref{sec:xpcomp}. We check them against the spectra in Section~\ref{sec:paramcheck}.}

\rev{We derive \teff\ from broad-band photometry.} We cross-match every star to the 2MASS catalog \citep{Skrutskie06} and combine the $JHK_s$ photometry with the \Gaia\ DR3 $G$, $G_{\rm BP}$, and $G_{\rm RP}$ magnitudes, correcting for the small foreground reddening. We take the reddening from the \Gaia\ DR3 \rev{General Stellar Parametrizer from Photometry (GSP-Phot)} $E(B_{\rm P}-R_{\rm P})$, converting it as $E(B-V)=E(B_{\rm P}-R_{\rm P})/1.3$. The sample is nearby and lightly reddened, with a median $E(B-V)\approx0.02$ and a maximum of about $0.1$. An error equal to the full median reddening changes the adopted \teff\ by at most $90$~K, within the adopted temperature uncertainty, and it enters the dwarfs and the giants alike, so it cannot produce a type-dependent effect. We compute $\teff$ from four color--temperature relations; the $(G-K_s)$, $(G_{\rm BP}-K_s)$, and $(G_{\rm RP}-K_s)$ calibrations of \citet{Mucciarelli21} and the $(J-K_s)$ calibration of \citet{GonzalezHernandez09}, all tied to interferometric and $(V-K_s)$ infrared-flux temperatures; and we also adopt their inverse-variance weighted mean. Each relation reproduces the solar temperature to within $30$--$55$~K, the four colors agree to a median of $90$~K per star, and the weighted mean has a typical internal precision of $30$~K; adding the color-to-color spread in quadrature gives an adopted $\sigma_{\teff}$ with a median of about $55$~K. For the error budget we adopt a more conservative $80$ to $120$~K per star (Section~\ref{sec:errors}), which allows for the systematic uncertainty of the color calibrations beyond this internal agreement.

We compute $\logg$ from the \Gaia\ DR3 parallax, the bolometric luminosity, the adopted $\teff$, and a mass of order $1\,M_\odot$, appropriate for old metal-rich disk stars. \rev{We assign the evolutionary state of each star from the nearest point on MIST isochrones \citep{Choi16,Dotter16} at $\feh=+0.4$, the median of our sample, with ages from $1$ to $14$~Gyr, in the \teff--\logg\ plane (Figure~\ref{fig:cmd} shows the same isochrones in the color-magnitude diagram). We call a star a dwarf if this point lies before the end of core hydrogen burning, a subgiant if it lies between that point and the base of the red giant branch, and a giant otherwise. This gives $33$ dwarfs, $7$ subgiants, and $16$ giants. Several warm dwarfs have $\logg<4$ because they lie near the turnoff of younger and more massive isochrones, which is why they sit above the solar-metallicity main sequence in Figure~\ref{fig:cmd}.} The adopted parameters are listed in Table~\ref{tab:results}, and the positions of the stars in the $\teff$--$\logg$ plane are shown in Figure~\ref{fig:kiel}.

\subsection{\rev{Line list and equivalent widths}}\label{sec:linelist}
We use the \ion{Fe}{1} and \ion{Fe}{2} line list of \citet{Trevisan11}. \rev{The oscillator strengths in this list are calibrated on the solar spectrum, so our scale is differential to the Sun.} From this list we keep the lines that fall in the region covered by the normalized spectra, $92$ lines in total (Table~\ref{tab:lines}). \rev{The lines we keep span excitation potentials from $0.9$ to $5.0$~eV and include the optical \ion{Fe}{2} lines of the list.}

\begin{deluxetable*}{cccc|cccc|cccc}
\tablecaption{Iron line list used for the equivalent-width analysis.\label{tab:lines}}
\tablehead{\colhead{$\lambda_{\rm air}$} & \colhead{Species} & \colhead{$\chi$} & \colhead{$\log gf$} & \colhead{$\lambda_{\rm air}$} & \colhead{Species} & \colhead{$\chi$} & \colhead{$\log gf$} & \colhead{$\lambda_{\rm air}$} & \colhead{Species} & \colhead{$\chi$} & \colhead{$\log gf$} \\
\colhead{(\AA)} & \colhead{} & \colhead{(eV)} & \colhead{} & \colhead{(\AA)} & \colhead{} & \colhead{(eV)} & \colhead{} & \colhead{(\AA)} & \colhead{} & \colhead{(eV)} & \colhead{}}
\startdata
5522.45 & \ion{Fe}{1} & 4.21 & $-1.49$ & 6079.01 & \ion{Fe}{1} & 4.65 & $-1.06$ & 6392.54 & \ion{Fe}{1} & 2.28 & $-4.06$ \\
5546.51 & \ion{Fe}{1} & 4.37 & $-1.18$ & 6082.71 & \ion{Fe}{1} & 2.22 & $-3.61$ & 6419.95 & \ion{Fe}{1} & 4.73 & $-0.38$ \\
5560.21 & \ion{Fe}{1} & 4.43 & $-1.14$ & 6084.11 & \ion{Fe}{2} & 3.20 & $-3.90$ & 6432.68 & \ion{Fe}{2} & 2.89 & $-3.60$ \\
5577.02 & \ion{Fe}{1} & 5.03 & $-1.61$ & 6093.64 & \ion{Fe}{1} & 4.61 & $-1.39$ & 6436.41 & \ion{Fe}{1} & 4.19 & $-2.38$ \\
5618.63 & \ion{Fe}{1} & 4.21 & $-1.39$ & 6113.33 & \ion{Fe}{2} & 3.22 & $-4.22$ & 6456.38 & \ion{Fe}{2} & 3.90 & $-2.17$ \\
5619.61 & \ion{Fe}{1} & 4.39 & $-1.51$ & 6127.90 & \ion{Fe}{1} & 4.14 & $-1.43$ & 6481.87 & \ion{Fe}{1} & 2.28 & $-2.98$ \\
5635.83 & \ion{Fe}{1} & 4.26 & $-1.65$ & 6149.26 & \ion{Fe}{2} & 3.89 & $-2.77$ & 6516.08 & \ion{Fe}{2} & 2.89 & $-3.33$ \\
5638.26 & \ion{Fe}{1} & 4.22 & $-0.83$ & 6151.62 & \ion{Fe}{1} & 2.18 & $-3.36$ & 6518.37 & \ion{Fe}{1} & 2.83 & $-2.58$ \\
5651.48 & \ion{Fe}{1} & 4.47 & $-1.86$ & 6157.73 & \ion{Fe}{1} & 4.07 & $-1.38$ & 6569.22 & \ion{Fe}{1} & 4.73 & $-0.61$ \\
5652.33 & \ion{Fe}{1} & 4.26 & $-1.81$ & 6159.38 & \ion{Fe}{1} & 4.61 & $-1.99$ & 6574.23 & \ion{Fe}{1} & 0.99 & $-4.99$ \\
5661.35 & \ion{Fe}{1} & 4.28 & $-1.91$ & 6165.37 & \ion{Fe}{1} & 4.14 & $-1.55$ & 6593.87 & \ion{Fe}{1} & 2.43 & $-2.44$ \\
5662.52 & \ion{Fe}{1} & 4.18 & $-0.65$ & 6170.51 & \ion{Fe}{1} & 4.79 & $-0.46$ & 6597.56 & \ion{Fe}{1} & 4.79 & $-1.01$ \\
5679.02 & \ion{Fe}{1} & 4.65 & $-0.84$ & 6173.34 & \ion{Fe}{1} & 2.22 & $-2.90$ & 6608.02 & \ion{Fe}{1} & 2.28 & $-4.03$ \\
5701.55 & \ion{Fe}{1} & 2.56 & $-2.22$ & 6180.20 & \ion{Fe}{1} & 2.73 & $-2.62$ & 6625.02 & \ion{Fe}{1} & 1.01 & $-5.31$ \\
5705.47 & \ion{Fe}{1} & 4.30 & $-1.52$ & 6187.99 & \ion{Fe}{1} & 3.94 & $-1.73$ & 6627.54 & \ion{Fe}{1} & 4.55 & $-1.53$ \\
5741.85 & \ion{Fe}{1} & 4.26 & $-1.71$ & 6200.31 & \ion{Fe}{1} & 2.61 & $-2.43$ & 6633.75 & \ion{Fe}{1} & 4.56 & $-0.80$ \\
5753.13 & \ion{Fe}{1} & 4.26 & $-0.77$ & 6213.43 & \ion{Fe}{1} & 2.22 & $-2.64$ & 6634.11 & \ion{Fe}{1} & 4.79 & $-1.08$ \\
5775.08 & \ion{Fe}{1} & 4.22 & $-1.16$ & 6219.28 & \ion{Fe}{1} & 2.20 & $-2.51$ & 6646.93 & \ion{Fe}{1} & 2.61 & $-4.05$ \\
5778.46 & \ion{Fe}{1} & 2.59 & $-3.57$ & 6220.78 & \ion{Fe}{1} & 3.88 & $-2.38$ & 6653.85 & \ion{Fe}{1} & 4.14 & $-2.52$ \\
5809.22 & \ion{Fe}{1} & 3.88 & $-1.73$ & 6226.74 & \ion{Fe}{1} & 3.88 & $-2.16$ & 6696.32 & \ion{Fe}{1} & 2.68 & $-3.73$ \\
5849.69 & \ion{Fe}{1} & 3.69 & $-3.14$ & 6229.23 & \ion{Fe}{1} & 2.84 & $-2.92$ & 6699.14 & \ion{Fe}{1} & 4.59 & $-2.15$ \\
5852.22 & \ion{Fe}{1} & 4.55 & $-1.24$ & 6240.65 & \ion{Fe}{1} & 2.22 & $-3.36$ & 6703.57 & \ion{Fe}{1} & 2.76 & $-3.06$ \\
5855.08 & \ion{Fe}{1} & 4.61 & $-1.60$ & 6247.56 & \ion{Fe}{2} & 3.89 & $-2.39$ & 6704.48 & \ion{Fe}{1} & 4.22 & $-2.60$ \\
5858.78 & \ion{Fe}{1} & 4.22 & $-2.17$ & 6265.13 & \ion{Fe}{1} & 2.18 & $-2.56$ & 6710.32 & \ion{Fe}{1} & 1.48 & $-4.90$ \\
5861.11 & \ion{Fe}{1} & 4.28 & $-2.51$ & 6270.23 & \ion{Fe}{1} & 2.86 & $-2.63$ & 6713.77 & \ion{Fe}{1} & 4.79 & $-1.49$ \\
5905.67 & \ion{Fe}{1} & 4.65 & $-0.87$ & 6297.79 & \ion{Fe}{1} & 2.22 & $-2.83$ & 6716.24 & \ion{Fe}{1} & 4.58 & $-1.78$ \\
5934.65 & \ion{Fe}{1} & 3.93 & $-1.26$ & 6311.50 & \ion{Fe}{1} & 2.83 & $-3.09$ & 6725.35 & \ion{Fe}{1} & 4.10 & $-2.29$ \\
5956.69 & \ion{Fe}{1} & 0.86 & $-4.59$ & 6315.81 & \ion{Fe}{1} & 4.07 & $-1.73$ & 6726.67 & \ion{Fe}{1} & 4.61 & $-1.09$ \\
5991.38 & \ion{Fe}{2} & 3.15 & $-3.63$ & 6322.69 & \ion{Fe}{1} & 2.59 & $-2.37$ & 6733.15 & \ion{Fe}{1} & 4.64 & $-1.47$ \\
6027.05 & \ion{Fe}{1} & 4.08 & $-1.23$ & 6369.46 & \ion{Fe}{2} & 2.89 & $-4.17$ & 6739.52 & \ion{Fe}{1} & 1.56 & $-4.96$ \\
6054.07 & \ion{Fe}{1} & 4.37 & $-2.31$ & 6380.74 & \ion{Fe}{1} & 4.19 & $-1.38$ & \nodata & \nodata & \nodata & \nodata \\
\enddata
\tablecomments{The complete list of 83 \ion{Fe}{1} and 9 \ion{Fe}{2} lines
($0.9<\chi<5.0$~eV), the solar-calibrated set of \citet{Trevisan11} restricted to the
region covered by our normalized spectra. The four columns are repeated three times across
the table and continue in reading order.}
\end{deluxetable*}

\begin{figure*}
\centering
\includegraphics[width=\textwidth]{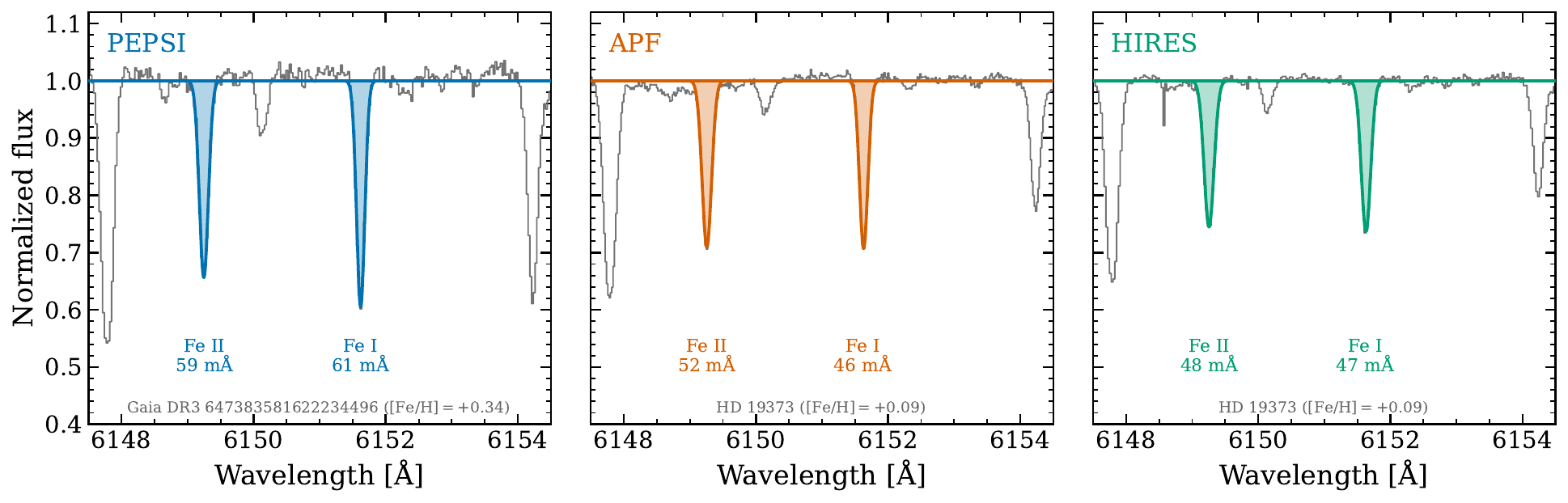}
\caption{Example of the equivalent-width measurement for the \ion{Fe}{2}~$\lambda6149.3$ and \ion{Fe}{1}~$\lambda6151.6$ lines on representative spectra from the three spectrographs. In each panel the dashed line marks the normalized continuum, the colored curve is the single-Gaussian fit, and the shaded area is the measured equivalent width. The APF and HIRES panels show the same star, the bright standard HD~19373, which is not part of the science sample but was observed with both spectrographs; because it is a single star, any difference between these two panels is purely instrumental, and the two equivalent widths agree to within a few m\AA. The PEPSI panel shows a more metal-rich science target, with correspondingly stronger lines.}
\label{fig:linefit}
\end{figure*}

We measure the equivalent width of each line in the rest-frame, normalized spectrum. For each line we set a local continuum from a linear fit to the upper envelope of the spectrum in two windows that lie between $0.45$ and $1.6$~\AA\ on either side of the line center, fit the core within $\pm0.4$~\AA\ of the center with a single Gaussian, and take the equivalent width as the area of that Gaussian (Figure~\ref{fig:linefit}). We keep a line only if it passes three tests: an equivalent width between about $10$ and \rev{$200$}~m\AA, a Gaussian width consistent with the instrumental width and the thermal width of iron, and a root-mean-square fit residual below $0.1$ in units of the normalized flux. \rev{We find that lines stronger than $120$~m\AA\ are affected by blends in their wings (Section~\ref{sec:valid}), so we measure them but exclude them from the abundance solution.} After these tests we have about $49$ to $80$ \ion{Fe}{1} lines and $3$ to $9$ \ion{Fe}{2} lines per star, of which $29$ to $80$ \ion{Fe}{1} lines lie below the $120$~m\AA\ \rev{limit} and enter the abundance solution. We check the equivalent widths against an independent measurement from a multi-Voigt profile fitter \citep{Egent} for a representative warm star. The two sets agree to a median of $1.5$~m\AA\ with a robust scatter of $2.7$~m\AA, and the strongest lines show no offset between the Gaussian and the Voigt area, so the single-Gaussian areas are not biased for these spectra. The line-by-line equivalent widths for all $62$ spectra, together with the adopted atomic data, are \rev{provided} as a machine-readable table (Table~\ref{tab:ews}).

\subsection{\rev{Abundances}}\label{sec:abund}
We convert the equivalent widths to abundances with \Korg\ \citep{Wheeler23,Wheeler24}, a one-dimensional LTE spectral-synthesis code that interpolates the MARCS grid of model atmospheres \citep{Gustafsson08}. We use the \texttt{ews\_to\_abundances} routine \rev{with model atmospheres at the star's \teff\ and \logg\ and at $\mh=+0.3$, close to the derived value}. We set the microturbulence $\vmic$ at fixed $\teff$ and $\logg$ by requiring the \ion{Fe}{1} abundance to show no trend with reduced equivalent width $\log({\rm EW}/\lambda)$\rev{, where EW is the equivalent width}, and we find the value of $\vmic$ that brings this slope to zero by bisection. Both the abundance and the \vmic\ solution use the \ion{Fe}{1} lines below the $120$~m\AA\ \rev{limit} (Section~\ref{sec:valid}).

\rev{If we set \vmic\ too low, the partly saturated lines return a higher abundance, and in these stars we find that an error in \vmic\ moves \feh\ at the $0.1$~dex level.} We take \feh\ as the median of the \ion{Fe}{1} line abundances, on a scale differential to the Sun with the adopted $A(\mathrm{Fe})_\odot=7.50$ \citep{Asplund09}. \rev{For the warm stars we find no residual trend of the \ion{Fe}{1} abundances with reduced equivalent width or excitation potential, and our \ion{Fe}{1} and \ion{Fe}{2} means agree. For the cool giants we find a small positive trend with excitation potential, which we examine in Section~\ref{sec:paramcheck}.}

\rev{We report abundances in LTE. For \ion{Fe}{1} at these parameters the non-LTE correction is between about $+0.03$ and $+0.05$~dex \citep{Amarsi16,Bergemann12,Lind12}, and we treat it as a systematic of at most a few hundredths of a dex.}

\subsection{Error budget}\label{sec:errors}
We combine four terms in quadrature for each star. The first is the standard error of the mean of the \ion{Fe}{1} line abundances, $\sigma/\sqrt{N}\approx0.02$ to $0.04$~dex. The second is the response of \feh\ to the adopted parameters, which we measure by repeating the analysis, including the $\vmic$ solution, at $\teff\pm\sigma_{\teff}$ and at $\logg\pm\sigma_{\logg}$, with $\sigma_{\teff}$ between $80$ and $120$~K and $\sigma_{\logg}\approx0.1$~dex. The third is a microturbulence term of about $0.04$~dex from the uncertainty in the reduced-equivalent-width slope. The fourth is a floor of about $0.03$~dex from the method and the continuum placement, set from the scatter of the benchmark results. The four terms give a total between about $0.05$ and $0.09$~dex (Table~\ref{tab:results}). The truncation of the line list is a further systematic term that we quote separately. \rev{Lowering the limit} from $120$ to $100$~m\AA\ \rev{changes the dwarfs by about $0.01$~dex, the subgiants by $0.04$~dex, and the giants by $0.09$~dex (medians)}, so the giant abundances carry an additional shared uncertainty of about $0.05$~dex from the choice of \rev{limit}.

\subsection{Validation}\label{sec:valid}
We apply the full procedure to two \Gaia\ FGK benchmark stars \citep{Jofre14,Heiter15}, with spectra from the library of \citet{BlancoCuaresma14}. These two stars span the parameter range of the sample: $\mu$~Leo, a K giant with $\teff=4474$~K, $\logg=2.51$, and reference $\feh=+0.25$, and $\mu$~Ara, a G dwarf with $\teff=5902$~K, $\logg=4.30$, and reference $\feh=+0.35$. We process the benchmark spectra through the same pipeline as the targets and obtain $\feh=+0.25$ for $\mu$~Leo and $\feh=+0.33$ for $\mu$~Ara, within $0.02$~dex of the reference values.

The strongest lines set the \rev{upper equivalent-width limit} of the line list. Our measurement tests accept lines up to about $200$~m\AA, but a test on $\mu$~Leo shows that the strongest lines bias the result. Truncating an extended $\mu$~Leo line list at $100$, $120$, and $140$~m\AA\ returns $\feh=+0.250$, $+0.231$, and $+0.172$ against the reference $+0.25$, and with no \rev{limit} it returns $+0.079$. The bias follows from blends in the wings of the strongest lines rather than from saturation, and \citet{Trevisan11}, whose oscillator strengths set our scale, restricted their own analysis to $20$ to $100$~m\AA\ for the same reason. We therefore base the abundances and the microturbulence on the lines below $120$~m\AA. We prefer $120$ to $100$~m\AA\ because four giant spectra would keep fewer than $30$ \ion{Fe}{1} lines under the tighter cut, while at $120$~m\AA\ every spectrum keeps at least $29$ lines and the balanced \vmic\ of the giants lies between $1.3$ and $1.9\kms$. Relative to \rev{a solution with no limit, the $120$~m\AA\ limit} raises the giants by about \rev{$0.14$}~dex, the subgiants by \rev{$0.06$}~dex, and the dwarfs by \rev{$0.03$}~dex, and the difference between the $120$ and $100$~m\AA\ \rev{limits} enters the error budget (Section~\ref{sec:errors}). The benchmark recovery above uses the same configuration, since neither benchmark line set contains lines above the \rev{limit}. We \rev{provide} the abundances obtained with no \rev{limit} and with \rev{limits} of $140$ and $100$~m\AA\ alongside the adopted values (see Section~\ref{sec:data_availability}).

We also test the 6 stars that were observed with multiple spectrographs. With the same parameters and the same common lines, the iron abundances from different spectrographs agree to between $0.01$ and $0.08$~dex, with a median difference of $0.03$~dex, which is set by a residual equivalent-width difference of a few m\AA\ between the reductions and by the standard error of the mean over the shared lines.

We also test the sample for spectroscopic binarity. We cross-correlate each spectrum against an \ion{Fe}{1} line mask. Every target shows a single cross-correlation peak at zero velocity with no second component, which is consistent with the ${\rm RUWE}<1.4$ astrometric selection, so all are single-lined. As a further check we inspected the \rev{Transiting Exoplanet Survey Satellite (TESS)} Quick Look Pipeline light curves \citep{Huang20} for the targets with TESS coverage. None shows eclipses or other photometric variability above the few-mmag level, so there is no evidence of binarity in the sample.

We also repeat the abundance step with a second synthesis code. We pass the same equivalent widths, line list, microturbulence, and MARCS model atmosphere for every spectrum to the {\tt MOOG} \rev{(November 2019 version)} \texttt{abfind} driver \citep{Sneden73}, so that any difference reflects the radiative transfer and curve-of-growth treatment alone. The two codes agree to within $0.04$~dex: {\tt MOOG} returns iron abundances a mean of $0.04$~dex below \Korg\ with a star-to-star scatter of $0.02$~dex, and the benchmarks show the same small shift, with {\tt MOOG} giving $\feh=+0.23$ for $\mu$~Leo and $+0.30$ for $\mu$~Ara against the \Korg\ values of $+0.25$ and $+0.33$. \citet{Griffith25} likewise find that \Korg\ returns iron abundances larger than {\tt MOOG} by $\lesssim0.05$~dex. The shift is constant and far smaller than both the metallicity range of the sample and the \Gaia-XP over-prediction, so neither the abundances nor the conclusion that the sample is metal-rich depends on the choice of code.

\subsection{Parameter verification} \label{sec:paramcheck}
\begin{figure*}
\centering
\includegraphics[width=\textwidth]{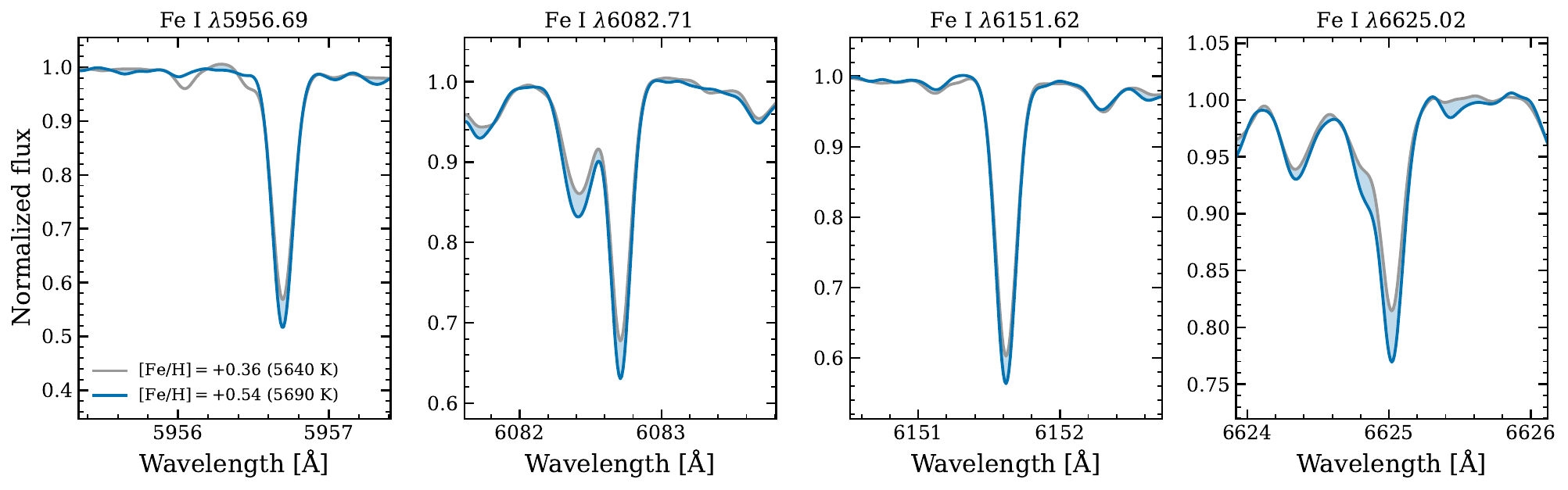}
\caption{Two dwarfs with two different metallicity values in the sample, with nearly identical parameters: $\feh = +0.36$ at $\teff \simeq 5640$~K and $\logg \simeq 4.30$ (grey) against $\feh = +0.54$ at $\teff \simeq 5690$~K and $\logg \simeq 4.35$ (blue), both observed with PEPSI. Each panel is centered on one of the measured \ion{Fe}{1} lines, with the spectra placed on a common wavelength and continuum scale. The shaded area marks the extra absorption of the more metal-rich star. Because the two stars have the same temperature and gravity to within the measurement uncertainties, the deeper lines of the blue spectrum reflect the higher iron abundance.}\label{fig:metseq}
\end{figure*}

We test the adopted parameters in three ways. First, we compare with the independent \Gaia\ GSP-Phot values: the adopted parallax \logg\ agrees with the GSP-Phot \logg\ to $\sim0.1$~dex, and our \teff\ agrees with the GSP-Phot \teff\ to a median absolute difference of $\sim70$~K for the 36 stars that have one. Second, we re-derived the \teff\ of each star from the spectrum alone by requiring the \ion{Fe}{1} abundance to show no trend with excitation potential. The warm stars show no significant excitation trend. The cool giants show a small positive slope, which would be removed only by a \teff\ more than $200$~K above the adopted value, a change larger than the calibration uncertainty. We return to this below. Third, we compare the \ion{Fe}{1} and \ion{Fe}{2} abundances, a check internal to the spectra. For the warmer stars the two agree within the \ion{Fe}{2} uncertainty, which is set by the small number of \ion{Fe}{2} lines. Most of the cool giants show \ion{Fe}{1} above \ion{Fe}{2}, with a mean difference of \rev{$+0.12$}~dex and individual values up to \rev{$0.33$}~dex, listed as the $\Delta_{\rm ion}$ column of Table~\ref{tab:results}, although a few show the reverse. This mean offset has the opposite sign to the \ion{Fe}{1} non-LTE correction, which is positive and would raise \ion{Fe}{1} further, so non-LTE cannot account for it. For the cool giants the second and third tests pull in opposite directions: the \teff\ increase that would flatten the excitation slope would drive \ion{Fe}{2} well below \ion{Fe}{1} and worsen the ionization balance. We \rev{interpret} this pair as a limitation of the cool-giant model atmospheres, where three-dimensional \citep{Collet07} and non-LTE \citep{Bergemann12,Lind12} effects and the few, blended \ion{Fe}{2} lines are hardest to model, rather than an error in the adopted temperature, and we keep the adopted photometric values. We therefore base \feh\ on the \ion{Fe}{1} lines, which are far more numerous and which satisfy the excitation balance for the warm stars.

The \ion{Fe}{1} abundance of the giants is in any case insensitive to the adopted \teff, changing by only about $0.03$~dex per $100$~K, and the cool benchmark giant $\mu$~Leo is recovered to $0.02$~dex with the same procedure (Section~\ref{sec:valid}). The over-prediction by \Gaia\ XP for the giants, which we present in Section~\ref{sec:xpcomp}, is therefore a property of the XP scale and not of our adopted temperatures.

\section{Results}\label{sec:results}

\subsection{Iron abundances and the most metal-rich stars}\label{sec:abundances}

\begin{figure*}
\centering
\includegraphics[width=\textwidth]{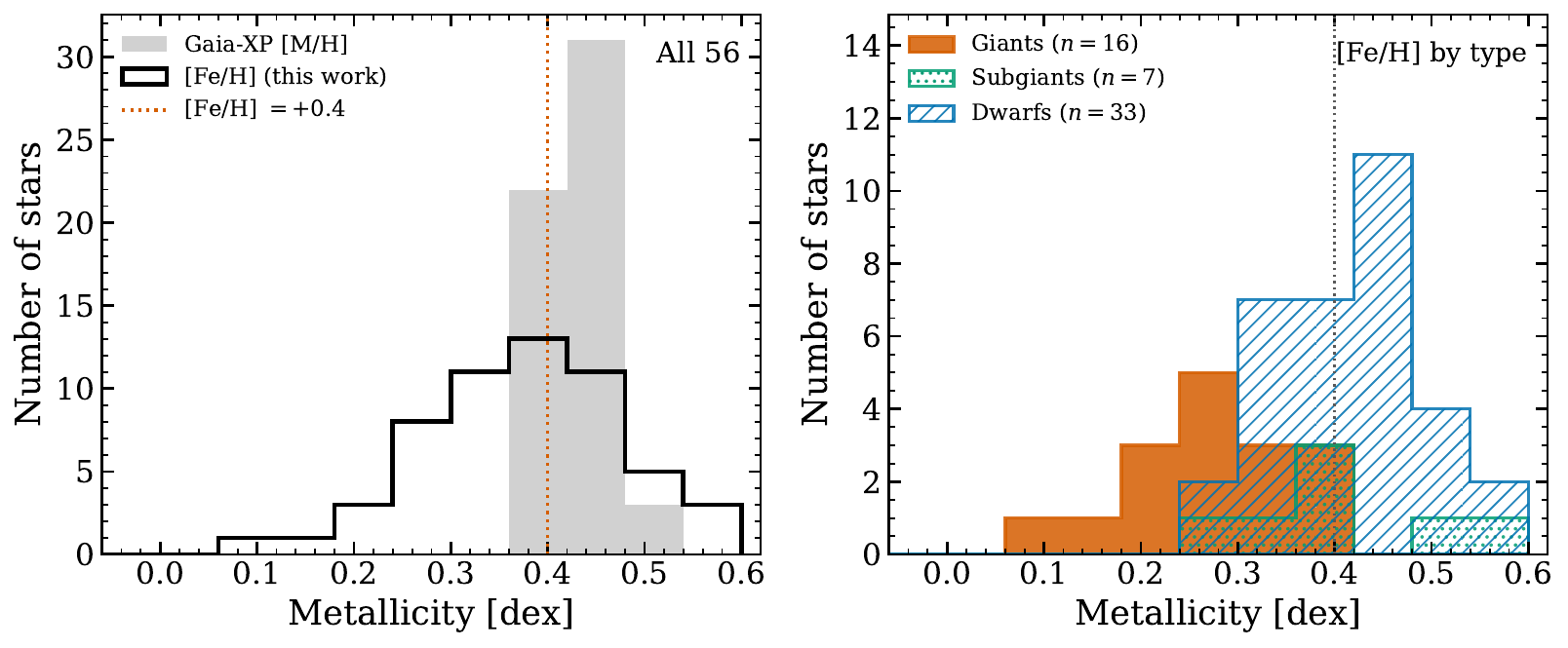}
\caption{\emph{Left:} metallicity distribution of all 56 PANTERA stars, comparing the \Gaia-XP values they were selected on (grey) with our spectroscopic \feh\ (black). The spectroscopic distribution is broader than the narrow XP band near $+0.4$, with a median at the same value and a tail toward lower abundances that belongs to the cool giants. A total of $25$ of the $56$ stars lie above \rev{$\feh=+0.4$} (dotted line). \emph{Right:} our \feh\ for the three evolutionary types (giants solid, subgiants dotted, dwarfs hatched). The dwarfs reach the \rev{highest metallicities}, the giants sit below their packed XP values, and the subgiants are intermediate.}
\label{fig:mdf}
\end{figure*}

\begin{figure}
\centering
\includegraphics[width=\columnwidth]{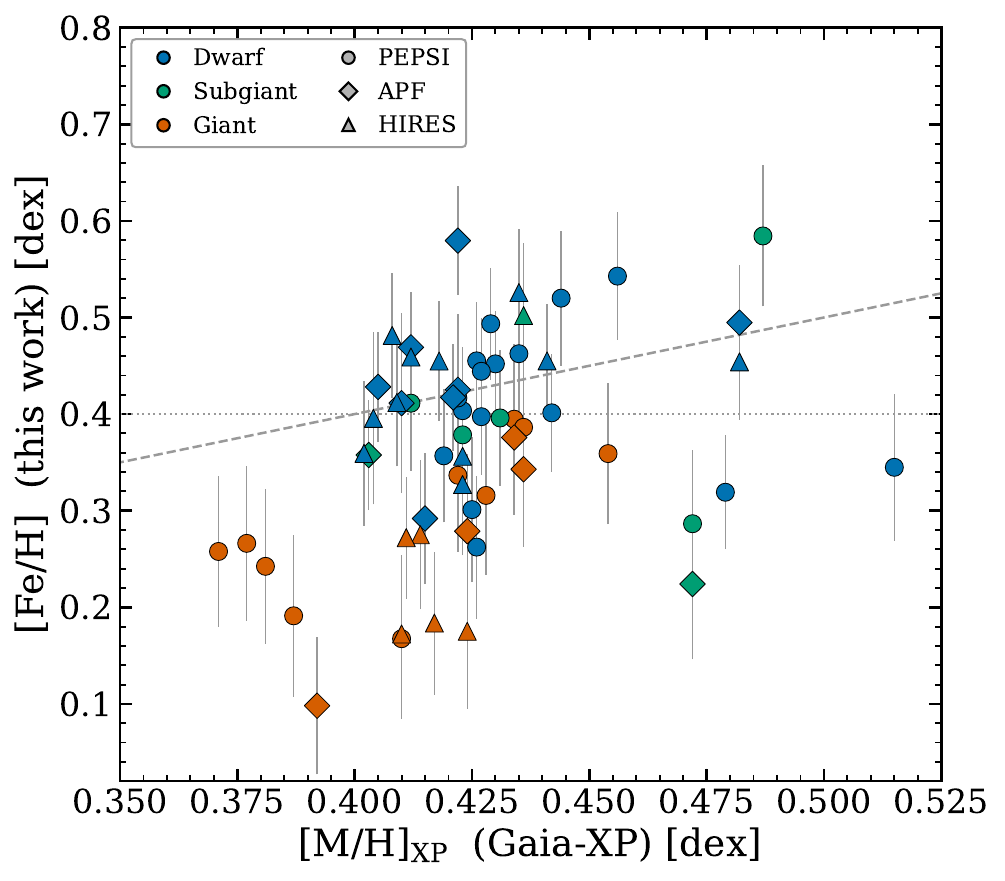}
\caption{Spectroscopic \feh\ vs. the \Gaia-XP \mh\ on which the targets were selected, for the 62 spectra of the 56 stars, with each instrument plotted separately. Color marks the evolutionary type and the symbol marks the instrument. The horizontal axis is zoomed on the narrow XP range of the sample. The giants lie below the one-to-one line (\rev{dashed}) while the dwarfs scatter about it, so XP over-predicts \feh\ for the evolved stars. The offset grows with evolutionary state, from a mean of \rev{$-0.01\pm0.01$}~dex for the dwarfs to \rev{$-0.05\pm0.04$}~dex for the subgiants and \rev{$-0.15\pm0.02$}~dex for the giants, while \Gaia\ XP compresses the sample into a narrow band near $+0.4$. The symbols of the three instruments overlap within each color, so the trend follows the stellar type and not the spectrograph. Because the parameters use no \Gaia\ BP/RP modeling (Section~\ref{sec:params}), the dwarf and giant offsets are placed on the same scale.}
\label{fig:results}
\end{figure}

\begin{figure}
\centering
\includegraphics[width=\columnwidth]{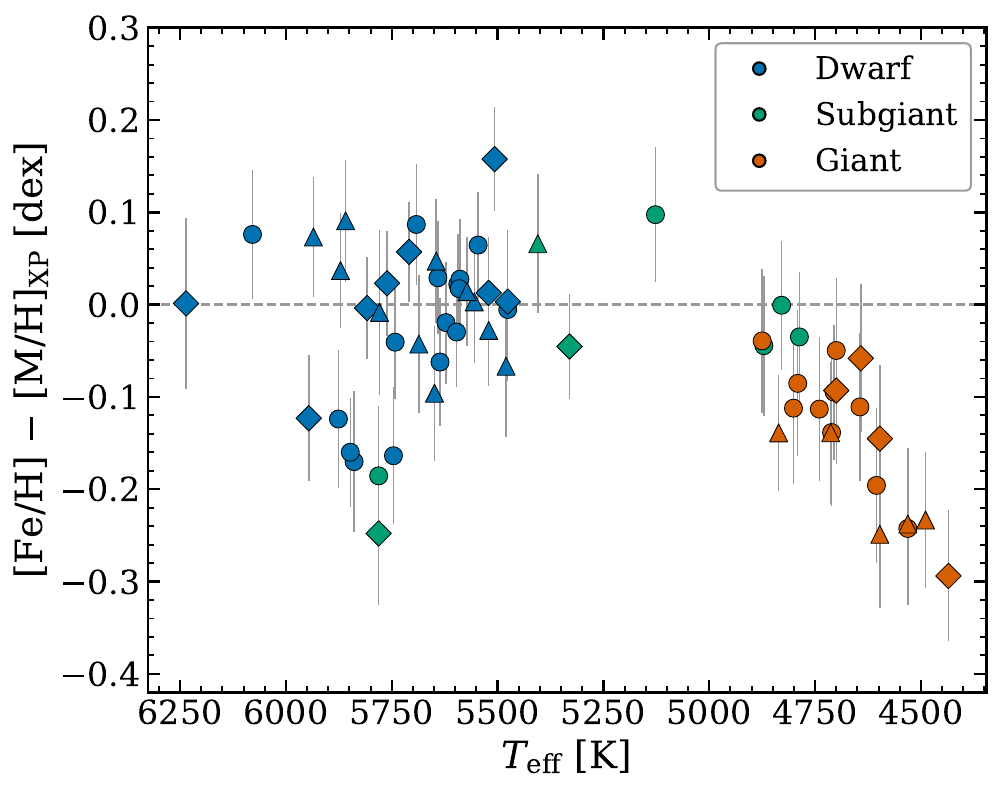}
\caption{Difference between the spectroscopic \feh\ and the \Gaia-XP \mh\ as a function of $\teff$, colored by evolutionary type and marked by instrument. The mean offset is \rev{$-0.01\pm0.01$}~dex for the dwarfs, \rev{$-0.05\pm0.04$}~dex for the subgiants, and \rev{$-0.15\pm0.02$}~dex for the cool giants, and the three instruments overlap within each type, so the over-prediction by \Gaia\ XP depends on the stellar type rather than on the spectrograph.}
\label{fig:offset}
\end{figure}

Table~\ref{tab:results} lists the adopted parameters and the iron abundances for the 56 stars. The abundances run from $\feh=+0.10$ for the coolest giants to $\feh=+0.58$, with a median of $+0.40$. The lowest values belong to the coolest giants, whose \Gaia-XP metallicities are the most overestimated (Section~\ref{sec:disc}). These stars set the low end of the range, while the sample as a whole stays metal-rich. Of the $56$ stars, $52$ \rev{have $\feh>+0.2$ and $25$ have $\feh>+0.4$. Most of the stars that XP placed near $+0.4$ are therefore SMR stars, and about half reach $+0.4$.} The $4$ stars at or below $+0.2$ are all cool giants, for which the \Gaia-XP metallicity is least reliable (Section~\ref{sec:disc}).

\rev{The most metal-rich giant in our sample has $\feh=+0.39$, while our dwarfs and subgiants reach $+0.58$. Part of this difference comes from our selection. We selected the giants at a lower XP threshold of $+0.35$, and we find that the XP values of the cool giants are too high by about \rev{$0.15$}~dex (Section~\ref{sec:xpcomp}), so for a giant we selected near $\mh_{\rm XP}\simeq+0.4$ we expect $\feh\simeq+0.25$. We do not expect mass loss or the first dredge-up to change the photospheric iron abundance. From our data alone we cannot tell whether giants above $\feh=+0.4$ are rare near the Sun or whether our XP selection misses them. \citet{Lu26} find a deficit of luminous metal-rich giants in \Gaia\ XP and spectroscopic survey data and suggest enhanced mass loss at high metallicity as one explanation. Our result is observational, and to test it directly we would need a sample of giants selected independently of the XP metallicities.}

We find that $25$ of the $56$ stars \rev{have $\feh>+0.4$}, with the most iron-rich stars reaching $\feh=+0.58$. Figure~\ref{fig:metseq} compares two dwarfs with nearly identical temperature and gravity with two different metallicity values, at $\feh=+0.36$ and $+0.54$, and the same instrument PEPSI. Every measured iron line of the more metal-rich star is visibly deeper. The two temperatures agree to within the uncertainty of the temperature scale, and the comparison does not rest on that agreement being exact: a common error in the scale moves both stars together, so repeating the analysis with both temperatures shifted by $120$~K changes the difference between the two abundances by less than $0.02$~dex. For these stars the excitation and reduced-equivalent-width balances are flat and the line-to-line scatter is small, so we are confident that the high abundances are well measured and are not an artifact of the XP selection. The most iron-rich stars, at $\feh=+0.58$, match the most metal-rich dwarfs of \citet{Trevisan11}, who report values up to $+0.58$ from an analysis restricted to \rev{lines weaker than about $100$~m\AA}, and lie well above the $\feh\simeq+0.4$ of the old open cluster NGC~6791 \citep{Brogaard12}.

These stars are useful targets for follow-up. They populate the high-metallicity end of the planet-metallicity relation, which is not well constrained above $\feh\simeq+0.3$, and their \Gaia\ astrometry together with \rev{the \Gaia\ radial velocities} can test whether the most iron-rich stars near the Sun migrated from the inner disk \citep{Sellwood02,Frankel18,Chen19,Nepal24}. \rev{We defer an analysis of their orbits, together with stellar ages, to the next paper in this series.}

The cool giants, though they are the least metal-rich stars in the sample on our scale, are the most relevant to mass loss on the red giant branch at high metallicity \citep{Hansen05,Tailo21,Miglio12}. A larger sample of confirmed metal-rich giants, with masses from asteroseismology, would give a direct test of the metallicity dependence of the mass-loss rate.

\subsection{Comparison with the \Gaia-XP metallicities}\label{sec:xpcomp}

\begin{figure*}
\centering
\includegraphics[width=\textwidth]{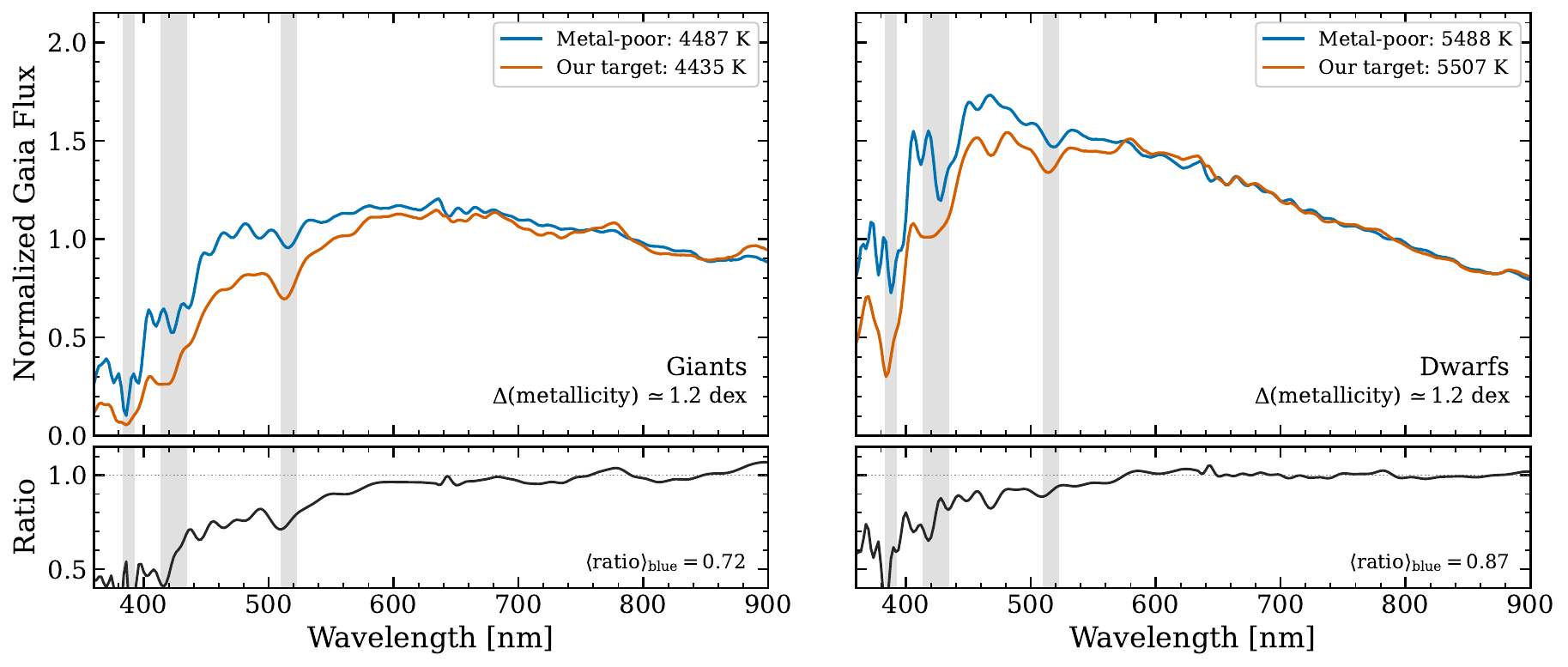}
\caption{\Gaia\ XP test of the blanketing origin of the over-prediction at fixed metallicity contrast. Each column compares a metal-rich target (red) with a metal-poor reference (blue) spanning $\sim1.2$~dex difference in metallicity between them, for giants ($\sim4460$~K, left) and dwarfs ($\sim5500$~K, right). The targets are a giant (Gaia DR3 4591824513402826112; $\feh=+0.10$, XP $\mh=+0.39$), for which the \ion{Fe}{1} and \ion{Fe}{2} abundances agree to $0.03$~dex, and our most iron-rich dwarf (Gaia DR3 2134294185689376512; $\feh=+0.58$, XP $\mh=+0.42$); the references are Gaia DR3 1918987234619955712 ($\mh\simeq-1.1$) and Gaia DR3 3402798414192387328 ($\mh\simeq-0.7$). Shaded bands mark CN ($\sim388$~nm), CH G-band ($\sim430$~nm), and MgH$+$Mg\,b ($\sim515$~nm); each spectrum is normalized in $760$--$820$~nm. Lower sub-panels: the metal-rich to metal-poor flux ratio, which over the blue ($390$--$520$~nm) falls to $0.72$ for the giants but only $0.87$ for the dwarfs.}
\label{fig:xpbands}
\end{figure*}

The high-resolution abundances are lower than the \Gaia-XP values that selected the sample. Figure~\ref{fig:mdf} compares the two distributions: the \Gaia-XP metallicities are packed into a narrow band near $+0.4$, while the spectroscopic abundances spread from $0.10$ to $+0.58$. Split by evolutionary type (Figure~\ref{fig:mdf}, right), the dwarfs reach the \rev{highest metallicities}, the cool giants sit well below their packed XP values, and the subgiants fall in between, a first view of the type-dependent over-prediction. Figure~\ref{fig:results} shows \feh\ against $\mh_{\rm XP}$ for the 56 stars. The offset between the two scales is not constant. It grows with the evolutionary state of the star, from a mean of \rev{$-0.01\pm0.01$}~dex for the dwarfs to \rev{$-0.05\pm0.04$}~dex for the subgiants and \rev{$-0.15\pm0.02$}~dex for the cool giants, where the quoted values are the mean and its standard error. Figure~\ref{fig:offset} shows this difference against $\teff$: it is small and consistent with zero for the warm dwarfs and grows smoothly toward the cool giants. The symbols of the three instruments overlap within each evolutionary class, so this trend follows the stellar type and not the spectrograph.

\section{Discussion}\label{sec:disc}

\subsection{The \Gaia-XP over-prediction and its origin}
\label{sec:xpover}

\begin{figure*}
\centering
\includegraphics[width=\textwidth]{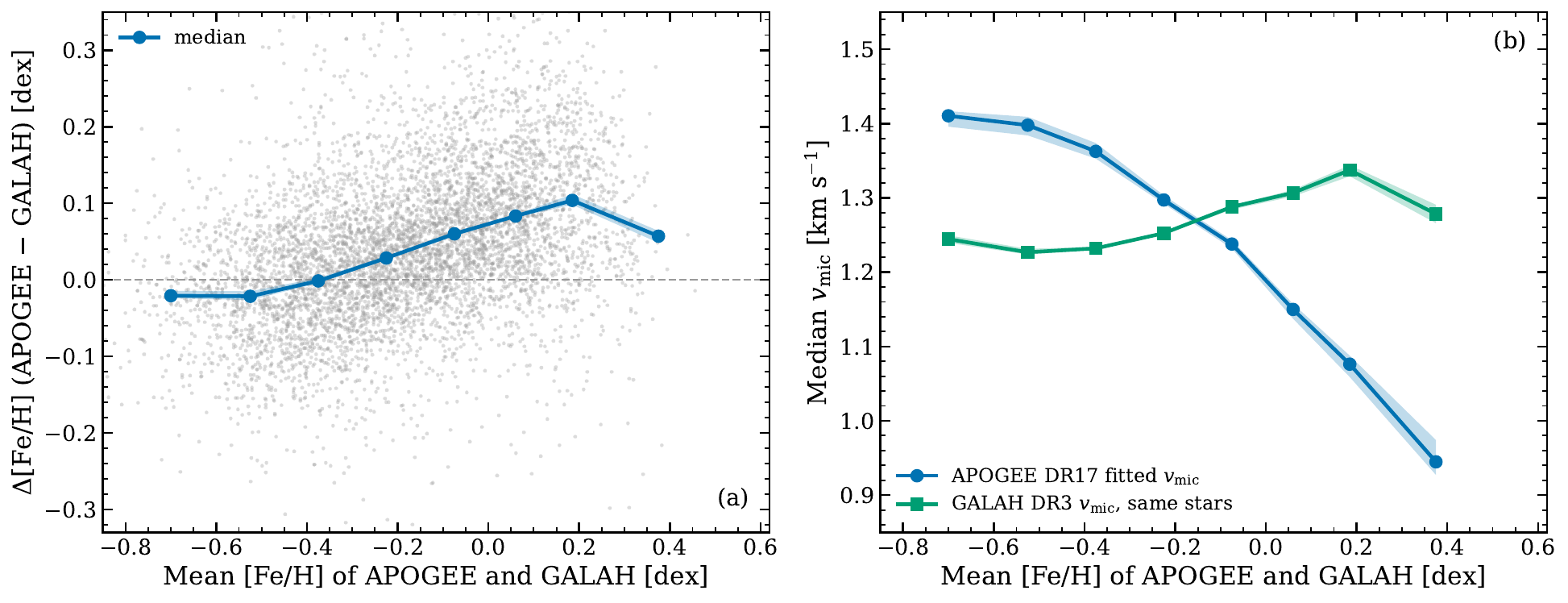}
\caption{The metal-rich giant scale, tested with the $6286$ cool giants ($4200<\teff<5000$~K, $2.0<\logg<3.6$) measured by both \rev{the Apache Point Observatory Galactic Evolution Experiment (APOGEE)} DR17 and \rev{the GALactic Archaeology with HERMES (GALAH)} DR3, matched by \Gaia\ source identifier and restricted to clean quality flags in both surveys. \emph{Left:} the difference between the APOGEE and GALAH iron abundances against the mean of the two, with grey points for the individual stars, the running median in blue, and bootstrap $68$ percent intervals as the band. The two surveys agree below solar metallicity, and APOGEE \rev{gives increasingly high values} toward the metal-rich end. \emph{Right:} the median microturbulence for the same stars. The APOGEE DR17 value, fitted as a free parameter, falls from $1.4$ to $0.9\kms$ as the metallicity rises, while the GALAH value, set by a calibrated relation, stays near $1.3\kms$. There is no physical reason for the turbulence of similar giants to decrease with metal content.}\label{fig:surveys}
\end{figure*}

\rev{The XP metallicities are higher than our \feh\ by an amount that grows from the dwarfs to the cool giants (Section~\ref{sec:xpcomp}). Because we take \teff\ and \logg\ from photometry and the parallax for all stars (Section~\ref{sec:params}), we measure the dwarf and giant offsets on the same scale.}

\rev{Part of this offset comes from a selection bias of about $0.1$~dex that is similar for all stellar types (Section~\ref{sec:dwarfscale}). It cancels in the difference between the types, so the growth of the offset from the dwarfs to the giants reflects a type-dependent error in the XP scale.}

\rev{We think the main reason is how the XP metallicity is measured. XP has a resolving power of only $R\sim100$, so the metallicity comes from the shape of the BP/RP continuum. In cool metal-rich giants this continuum is depressed by the CN, CH, and MgH bands and the dense metal lines, mostly in the blue, and the XP training set holds few such stars \citep{Andrae23,Zhang23}. Figure~\ref{fig:xpbands} shows this at a fixed metallicity contrast of $\sim1.2$~dex. Averaged over $390$--$520$~nm, the ratio of the metal-rich to the metal-poor spectrum is $0.72$ for the giants but $0.87$ for the dwarfs, so the blue blanketing is much stronger in the cool giants.}

\rev{The over-prediction grows smoothly with \teff\ and \logg\ across the sample. This matters for maps built from XP metallicities, for example the metal-rich ``knot'' of giants at the Galactic center that \citet{Rix24} select at $\mh_{\rm XP}>+0.5$, the regime in which we measure the largest over-prediction. We plan to derive a \teff- and \logg-dependent correction in future work. Our sample covers only the metal-rich end, so these results need not apply at lower metallicity.}

\begin{figure}
\centering
\includegraphics[width=\columnwidth]{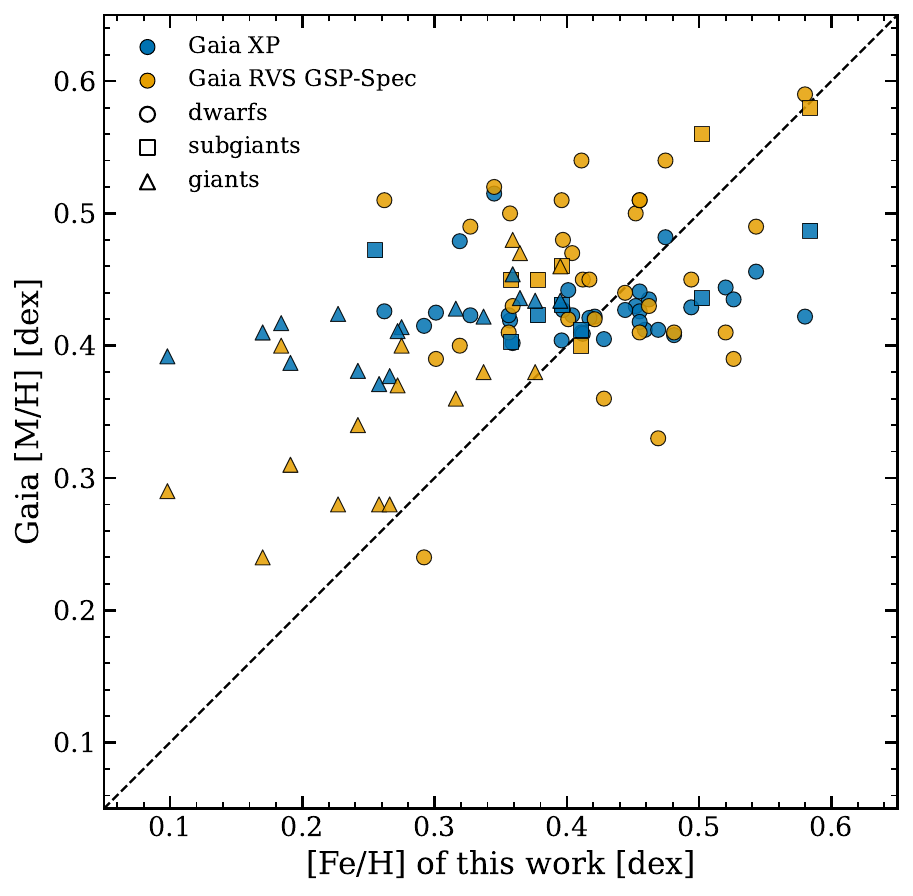}
\caption{The \Gaia-XP (blue) and \Gaia\ RVS GSP-Spec (orange) metallicities of the PANTERA stars against our spectroscopic \feh. Circles, squares, and triangles mark dwarfs, subgiants, and giants. For the dwarfs both products scatter about the one-to-one line, while for the cool giants the XP values pack near $+0.4$ and the RVS values fall between the XP values and our scale.}\label{fig:rvs}
\end{figure}

The giant offset points to the XP scale rather than to our adopted giant abundances. The \feh\ we derive for the cool giants depends on the treatment of the strongest lines at about the $0.15$~dex level (Section~\ref{sec:valid}). An equivalent-width solution \rev{with no limit} returns giant abundances lower by about $0.14$~dex, and a full-spectrum synthesis fit over the red \ion{Fe}{1} lines returns lower values still, with $\feh=+0.05$ for $\mu$~Leo against the reference $+0.25$, because those lines are saturated and blended in cool metal-rich giants. The equivalent-width analysis \rev{with the $120$~m\AA\ limit} is the configuration that recovers the benchmarks (Section~\ref{sec:valid}), and the alternatives would lower the giant abundances and enlarge the XP over-prediction, so the \rev{$0.15$}~dex giant offset is the smallest of the available \rev{estimates} and is not an artifact of our analysis.

\subsection{The metal-rich giant scale in the large surveys}\label{sec:surveys}

The blanketing explanation above concerns the XP spectra themselves. A second, related question is where the XP metallicity scale comes from. The XGBoost values we selected on are trained on \rev{Apache Point Observatory Galactic Evolution Experiment (APOGEE)} DR17 labels \citep{Andrae23}, so any systematic error in the APOGEE metallicities of cool metal-rich giants would be inherited by the XP values for the same kind of stars. We test the APOGEE scale against \rev{GALactic Archaeology with HERMES (GALAH)} DR3 \citep{Buder21}, an optical survey with an independent pipeline, using the $6286$ cool giants of similar \logg\ and \teff\ cut as ours, measured by both surveys (Figure~\ref{fig:surveys}). The two surveys agree below solar metallicity, but \rev{the APOGEE values are higher} with increasing metallicity, by about $+0.1$~dex in the mean at $\mh\simeq+0.3$ and by $+0.19$~dex for the giants that APOGEE places above $+0.3$ at $\teff<4700$~K. The likely reason is the microturbulence. APOGEE DR17 fits \vmic\ as a free parameter of its global solution, a change from the surface-gravity relation of the earlier releases \citep{GarciaPerez16}, and the fitted values fall to unphysically low $0.9$--$1.1\kms$ for the metal-rich cool giants, while the calibrated GALAH values for the same stars stay near $1.3 \kms$ (Figure~\ref{fig:surveys}, right). GALAH carries its own systematics for cool metal-rich giants, so this comparison \rev{points to a possible calibration error} on the APOGEE side rather than a statement that the GALAH scale is the truth. The internal APOGEE evidence below does not rest on GALAH. As Section~\ref{sec:abund} describes, a \vmic\ set too low forces the saturated iron lines to return a higher abundance. For the benchmark giant $\mu$~Leo, adopting $\vmic=1.0\kms$ in place of the balanced $1.5\kms$ raises the recovered \feh\ from $+0.25$ to $+0.49$. Consistent with an APOGEE-side effect, 12 of our targets have APOGEE DR17 abundances. Eleven of the twelve agree with our \feh\ to a mean of $+0.01$~dex, all within $0.11$~dex, while the one cool giant in the overlap ($\teff\approx4490$~K) is higher in APOGEE by $0.27$~dex even though APOGEE adopts an equal or cooler temperature for it, exactly the regime where the fitted \vmic\ falls lowest.

The XP metallicities follow the APOGEE values almost exactly for these giants (median difference $+0.005$~dex with a scatter of $0.05$~dex), as expected for a data-driven method applied where the training labels carry the signal. The \Gaia\ RVS \rev{General Stellar Parametrizer from Spectroscopy (GSP-Spec)} values \citep{RecioBlanco23}, which come from a model-based pipeline that does not train on APOGEE, show about half of the metal-rich rise. All 56 of our targets have GSP-Spec values, and for the cool giants they fall between our scale and the XP values (Figure~\ref{fig:rvs}). We conclude that the XP over-prediction of the cool metal-rich giants likely combines the blue blanketing of the XP spectra with systematics inherited from the training labels. Separating the two contributions requires an XP metallicity trained on an independent reference. With a larger confirmed sample that fills the \teff--\logg\ plane, we plan to provide an applicable correction to the existing photometric and survey metallicity catalogs at the metal-rich end in future work.

\subsection{Selection bias and the dwarf scale}\label{sec:dwarfscale}

The measured offset includes a selection bias that affects all stellar types. The XP metallicities have an uncertainty near $0.1$~dex, so a cut at a high XP value preferentially admits stars whose XP metallicity is scattered upward. The spectroscopic abundances of the selected sample then fall below the XP values even where the two scales agree in the mean \citep{Eddington13}. A forward model of this effect, applying an XP scatter $\sigma_{\rm XP}\simeq0.1$~dex to the steeply falling metal-rich metallicity function (e-folding $\simeq0.1$~dex) with our selection cuts, gives a minimum offset of about $0.1$~dex even if the XP scale is unbiased. This sets a floor on the offset we measure. The floor scales as $\sigma_{\rm XP}^2$, so it is comparable for all stellar types unless the XP scatter rises steeply toward the cool giants, and it sets a floor on the offset rather than a trend with stellar type. For the cool giants the measured \rev{$0.15$}~dex offset exceeds this floor, and the floor cancels in the difference between the types, so the growth of the offset from the dwarfs to the giants is free of it. For the dwarfs the measured offset (\rev{$-0.01$}~dex) sits above the floor prediction: either the floor overestimates the selection bias for our sample, or the XP values of the dwarfs are intrinsically low by about the size of the floor and the two effects cancel. The re-analysis of the \citet{Trevisan11} dwarfs below points to the second \rev{interpretation}. Either way the near-zero dwarf offset should not be \rev{interpreted} as a confirmation of the XP dwarf scale.

The metal-rich dwarfs of \citet{Trevisan11} lie above their \Gaia-XP values, so for that sample XP under-predicts the metallicity by up to about $0.25$~dex, an offset of the opposite sign to our near-zero dwarf result. To check that this is not a scale difference between our analysis and theirs, we re-analyzed the $32$ out of $71$ \citet{Trevisan11} stars that have public archival spectra (HARPS, UVES, FEROS, and ESPRESSO) with our pipeline, and we recovered their iron abundances to $0.05$~dex, in line with the $0.02$~dex recovery of the \Gaia\ benchmarks. Our scale therefore agrees with theirs, and the difference between the two dwarf offsets is one of sample selection rather than of measurement. Because we select our dwarfs at high XP, the selection bias pushes our measured dwarf offset downward and can mask an intrinsic under-prediction. The near-zero value we measure is consistent both with a genuinely accurate XP dwarf scale and with a modest under-prediction hidden by selection. We therefore do not \rev{interpret} the dwarf agreement as a measurement of the XP accuracy, and a sample spanning a wider and lower XP range would be needed to pin the dwarf relation directly.

\section{Conclusions \& Future Work}\label{sec:conclusions}

We have presented the first result from PANTERA, a program to survey and understand the most metal-rich stars in the solar neighborhood. This first paper begins with a sample selected from the \Gaia-XP metallicities. We report iron abundances for 56 stars observed with PEPSI, the APF, and HIRES. Our main results are as follows.

\begin{enumerate}
\item We measure \feh\ from an equivalent-width analysis with fixed photometric temperatures and parallax gravities, using the \ion{Fe}{1} lines below a $120$~m\AA\ \rev{upper limit} set by the benchmark tests. We recover the $\mu$~Leo and $\mu$~Ara benchmark metallicities to within $0.02$~dex, the stars observed with more than one spectrograph agree to within $0.08$~dex, and a second synthesis code and an independent equivalent-width measurement give the same result.

\item The sample is metal-rich, with abundances from $+0.10$ to $+0.58$ and a median of $+0.40$. 25 of the 56 stars \rev{have $\feh>+0.4$}, with the most iron-rich stars reaching $\feh=+0.58$.

\item The \Gaia-XP metallicities are higher than our spectroscopic \feh\ by a margin that grows with evolutionary state, from \rev{$-0.01\pm0.01$}~dex for the warm dwarfs to \rev{$-0.05\pm0.04$}~dex for the subgiants and \rev{$-0.15\pm0.02$}~dex for the cool giants. The offset follows \teff\ and \logg\ and not the instrument, and a roughly constant selection bias cannot account for the trend, so the cool-giant over-prediction is a property of the XP scale.

\item We attribute the discrepancy to two effects that act together. The first is the strong blue line blanketing of cool metal-rich giants, which lowers the BP/RP continuum that the XP metallicity is read from and which is difficult to reproduce in the model atmospheres behind a data-driven metallicity. The second is the APOGEE DR17 labels on which the XP metallicities are trained, which carry a related bias for the same stars that we trace to the freely fitted microturbulence of the DR17 pipeline. We do not separate the two contributions here. The XP metallicity at the metal-rich end therefore depends on stellar type, and a correction that depends on \teff\ and \logg\ would be needed before the XP metal-rich giants are used for a quantitative study.

\end{enumerate}

\section*{Data Availability}\label{sec:data_availability}
The equivalent widths of all $62$ spectra with their atomic data (Table~\ref{tab:ews}), the adopted parameters and iron abundances (Table~\ref{tab:results}), the iron abundances and microturbulences obtained with no equivalent-width \rev{limit} and with \rev{limits} of $140$ and $100$~m\AA\ alongside the adopted $120$~m\AA\ values (Section~\ref{sec:valid}), \rev{and the ADQL selection queries (Section~\ref{sec:select}),} are available at \rev{\url{https://github.com/panterasurvey/pantera1}}. The reduced spectra are available on request.

\section*{Acknowledgments}

We thank Lucy Lu for useful discussions about target selection. We thank Yuan-Sen Ting for comments and suggestions.

SMS is supported by the Distinguished University Fellowship awarded by The Ohio State University. DMR is supported by NASA Hubble Fellowship grant HST-HF2-51588.001-A awarded by the Space Telescope Science Institute, which is operated by the Association of Universities for Research in Astronomy, Inc., for NASA, under contract NAS5-26555. KZS is partially supported by NSF grants AST-2307385 and AST-2407206.

The LBT is an international collaboration among institutions in the United States and Europe. At the time data were acquired for this research, LBT Corporation Members were the University of Arizona on behalf of the Arizona Board of Regents; Istituto Nazionale di Astrofisica, Italy; and The Ohio State University, representing The Ohio State University, University of Notre Dame, University of Minnesota, and University of Virginia.  This research used the facilities of the Italian Center for Astronomical Archives (IA2) operated by INAF at the Astronomical Observatory of Trieste.  Observations have benefited from the use of ALTA Center (alta.arcetri.inaf.it) forecasts performed with the Astro-Meso-Nh model. Initialization data of the ALTA automatic forecast system come from the General Circulation Model (HRES) of the European Centre for Medium Range Weather Forecasts.

The data presented herein were obtained in part at the W.~M.~Keck Observatory, which is operated as a scientific partnership among the California Institute of Technology, the University of California, and the National Aeronautics and Space Administration. The Observatory was made possible by the generous financial support of the W.~M.~Keck Foundation. The authors wish to recognize and acknowledge the very significant cultural role and reverence that the summit of Maunakea has always had within the indigenous Hawaiian community, and we are most fortunate to have the opportunity to conduct observations from this mountain.

This work also made use of observations obtained with the Automated Planet Finder at Lick Observatory, operated by the University of California Observatories.

This work has made use of data from the European Space Agency (ESA) mission \textit{Gaia} (\url{https://www.cosmos.esa.int/gaia}), processed by the \textit{Gaia} Data Processing and Analysis Consortium (DPAC, \url{https://www.cosmos.esa.int/web/gaia/dpac/consortium}). Funding for the DPAC has been provided by national institutions, in particular the institutions participating in the \textit{Gaia} Multilateral Agreement.

This publication makes use of data products from the Two Micron All Sky Survey (2MASS), which is a joint project of the University of Massachusetts and the Infrared Processing and Analysis Center/California Institute of Technology, funded by the National Aeronautics and Space Administration and the National Science Foundation.

We used Claude Code (Anthropic) for coding assistance to automate our existing, human-written abundance-measurement pipeline and to improve the presentation of the figures. We read and ran all AI-written code ourselves. We used Egent \citep{Egent}, an LLM-based agent, only as an independent check of our equivalent widths.

\facility{LBT (PEPSI), APF (Levy), Keck:I (HIRES), TESS}
\software{Korg \citep{Wheeler23}, MOOG \citep{Sneden73}, Egent \citep{Egent}, lightkurve \citep{lightkurve}, astropy \citep{astropy}, numpy, scipy, matplotlib}

\begin{deluxetable*}{lllcccc}
\tablecaption{Line-by-line equivalent-width measurements.\label{tab:ews}}
\tablewidth{0pt}
\tablehead{
\colhead{Gaia DR3} & \colhead{Inst.} & \colhead{Species} &
\colhead{$\lambda_{\rm air}$} & \colhead{$\chi$} &
\colhead{$\log gf$} & \colhead{EW} \\
\colhead{} & \colhead{} & \colhead{} &
\colhead{(\AA)} & \colhead{(eV)} & \colhead{} & \colhead{(m\AA)}
}
\startdata
647383581622234496 & PEPSI & \ion{Fe}{1} & 5522.45 & 4.21 & $-1.49$ & 57.8 \\
647383581622234496 & PEPSI & \ion{Fe}{1} & 5546.51 & 4.37 & $-1.18$ & 72.5 \\
647383581622234496 & PEPSI & \ion{Fe}{1} & 5560.21 & 4.43 & $-1.14$ & 69.4 \\
647383581622234496 & PEPSI & \ion{Fe}{1} & 5618.63 & 4.21 & $-1.39$ & 67.5 \\
647383581622234496 & PEPSI & \ion{Fe}{2} & 6084.11 & 3.20 & $-3.90$ & 45.4 \\
1291436403222476928 & APF & \ion{Fe}{1} & 5522.45 & 4.21 & $-1.49$ & 82.6 \\
1291436403222476928 & APF & \ion{Fe}{1} & 5546.51 & 4.37 & $-1.18$ & 135.4 \\
1291436403222476928 & APF & \ion{Fe}{1} & 5560.21 & 4.43 & $-1.14$ & 82.6 \\
1291436403222476928 & APF & \ion{Fe}{2} & 6149.26 & 3.89 & $-2.77$ & 52.5 \\
1172246044236127744 & HIRES & \ion{Fe}{1} & 5522.45 & 4.21 & $-1.49$ & 77.2 \\
1172246044236127744 & HIRES & \ion{Fe}{1} & 5546.51 & 4.37 & $-1.18$ & 111.8 \\
1172246044236127744 & HIRES & \ion{Fe}{1} & 5560.21 & 4.43 & $-1.14$ & 75.8 \\
1172246044236127744 & HIRES & \ion{Fe}{2} & 6084.11 & 3.20 & $-3.90$ & 35.2 \\
\enddata
\tablecomments{The adopted wavelengths, excitation potentials $\chi$, and
oscillator strengths $\log gf$ are from the differential line list of
\citet{Trevisan11}. Table~\ref{tab:ews} is published in its entirety, for all
$62$ spectra ($4{,}700$ lines), at \rev{\url{https://github.com/panterasurvey/pantera1}}. A portion is
shown here for guidance regarding its form and content.}
\end{deluxetable*}

\begin{deluxetable*}{llcccccccc}
\tabletypesize{\footnotesize}
\tablecaption{PANTERA stars: adopted parameters and \Korg\ EW-equilibrium iron abundances. Effective temperatures are on the homogeneous infrared-flux scale (Section~\ref{sec:params}) and gravities from the \Gaia\ parallax. The microturbulence rises with evolutionary state, with median values of \rev{$1.28$, $1.34$, and $1.56\kms$} for the dwarfs, subgiants, and giants. A representative subset is shown here; the full table for all 62 spectra of 56 stars is published in the machine-readable form.\label{tab:results}}
\tablehead{\colhead{\Gaia\ DR3 source\_id} & \colhead{Inst.} & \colhead{\teff} & \colhead{\logg} & \colhead{\vmic} & \colhead{\feh} & \colhead{[\ion{Fe}{2}/H]} & \colhead{$\Delta_{\rm ion}$} & \colhead{$N_{\rm Fe}$} & \colhead{\mh$_{\rm XP}$} \\ \colhead{} & \colhead{} & \colhead{(K)} & \colhead{(dex)} & \colhead{(\kms)} & \colhead{(dex)} & \colhead{(dex)} & \colhead{(dex)} & \colhead{} & \colhead{(dex)}}
\startdata
4591824513402826112 & APF & 4435 & 2.60 & 1.86 & $+0.10\pm0.07$ & $+0.13$ & $-0.03$ & 32 & $+0.39$ \\
1332825044549520640 & HIRES & 4489 & 2.88 & 1.66 & $+0.18\pm0.07$ & $+0.56$ & $-0.38$ & 40 & $+0.42$ \\
4547137062312951680 & HIRES & 4531 & 2.51 & 1.68 & $+0.17\pm0.08$ & $+0.02$ & $+0.15$ & 37 & $+0.41$ \\
4547137062312951680 & PEPSI & 4531 & 2.51 & 1.74 & $+0.17\pm0.08$ & $-0.14$ & $+0.31$ & 37 & $+0.41$ \\
1615356205256881920 & PEPSI & 4605 & 2.62 & 1.71 & $+0.19\pm0.08$ & $-0.08$ & $+0.27$ & 29 & $+0.39$ \\
1373298170645854208 & PEPSI & 4740 & 3.14 & 1.54 & $+0.26\pm0.08$ & $-0.03$ & $+0.29$ & 38 & $+0.37$ \\
4285843204916407296 & APF & 4700 & 3.63 & 1.56 & $+0.34\pm0.08$ & $+0.46$ & $-0.12$ & 52 & $+0.44$ \\
4285843204916407296 & PEPSI & 4700 & 3.63 & 1.30 & $+0.39\pm0.08$ & $+0.44$ & $-0.05$ & 55 & $+0.44$ \\
3825993705211972864 & PEPSI & 4705 & 3.52 & 1.46 & $+0.36\pm0.07$ & $+0.37$ & $-0.01$ & 38 & $+0.45$ \\
744784953040324864 & PEPSI & 4787 & 3.72 & 1.27 & $+0.40\pm0.07$ & $+0.27$ & $+0.12$ & 47 & $+0.43$ \\
1835520283318364800 & APF & 5330 & 4.00 & 1.40 & $+0.36\pm0.06$ & $+0.34$ & $+0.02$ & 54 & $+0.40$ \\
2250537578633588352 & HIRES & 5405 & 4.01 & 1.50 & $+0.50\pm0.07$ & $+0.23$ & $+0.28$ & 59 & $+0.44$ \\
4311008797037316224 & APF & 5476 & 4.11 & 1.27 & $+0.42\pm0.08$ & $+0.50$ & $-0.08$ & 69 & $+0.42$ \\
4311008797037316224 & PEPSI & 5476 & 4.11 & 1.20 & $+0.42\pm0.08$ & $+0.40$ & $+0.02$ & 75 & $+0.42$ \\
\enddata
\tablecomments{This table is published in its entirety, for all $62$ spectra of $56$ stars, at \rev{\url{https://github.com/panterasurvey/pantera1}}. A portion is shown here for guidance regarding its form and content.}
\end{deluxetable*}

\bibliographystyle{mnras}
\bibliography{refs}

\end{document}